# Discovery of new antiviral agents through artificial intelligence: *in vitro* and *in vivo* results

Roza Izmailyan[1], Mher Matevosyan[2], Hamlet Khachatryan[1,2], Anastasiya Shavina[1,2], Smbat Gevorgyan[1,2], Artur Ghazaryan[1], Irina Tirosyan[2], Yeva Gabrielyan[2], Marusya Ayvazyan[2], Boris Martirosyan[2], Vardan Harutyunyan[2], Hovakim Zakaryan[1,2,*]

[1]Laboratory of Antiviral Drug Discovery, Institute of Molecular Biology of NAS, Hasratyan 7, 0014, Yerevan, Armenia.

[2]Denovo Sciences Inc., 0060, Yerevan, Armenia.

*Corresponding author: Hovakim Zakaryan, e-mail: h_zakaryan@mb.sci.am

## Abstract

In this research, we employed a deep reinforcement learning (RL)-based molecule design platform to generate a diverse set of compounds targeting the neuraminidase (NA) of influenza A and B viruses. A total of 60291 compounds were generated, of which 86.5% displayed superior physicochemical properties compared to oseltamivir. After narrowing down the selection through computational filters, nine compounds with non-sialic acid-like structures were selected for *in vitro* experiments. We identified two compounds, DS-22-inf-009 and DS-22-inf-021 that effectively inhibited the NAs of both influenza A and B viruses (IAV and IBV), including H275Y mutant strains at low micromolar concentrations. Molecular dynamics simulations revealed a similar pattern of interaction with amino acid residues as oseltamivir. In cell-based assays, DS-22-inf-009 and DS-22-inf-021 inhibited IAV and IBV in a dose-dependent manner with $EC_{50}$ values ranging from 0.29 µM to 2.31 µM. Furthermore, animal experiments showed that both DS-22-inf-009 and DS-22-inf-021 exerted antiviral activity in mice, conferring 65% and 85% protection from IAV (H1N1 pdm09), and 65% and 100% protection from IBV (Yamagata lineage), respectively. Thus, these findings demonstrate the potential of RL to generate compounds with promising antiviral properties.

## Introduction

The conventional approaches to antiviral research and vaccine development face the challenge of time-consuming processes. The critical importance of swift discovery of novel antiviral agents cannot be overstated, particularly in the context of emerging viruses that demand rapid response measures. While vaccines play a pivotal role in controlling the spread of viruses, their development and manufacturing timelines often span months or years. As a result, antiviral drugs stand as the frontline defense against evolving viral threats, offering crucial management of severe infections and mitigating morbidity and mortality until effective vaccines developed. In this landscape of evolving viral threats, the integration of cutting-edge technologies into antiviral drug discovery becomes an essential step.

Machine learning (ML) is an emerging field that has positively impacted the drug discovery process (Vamathevan et al., 2019; Patel et al., 2020). Different ML algorithms are currently applied in various aspects of drug discovery, such as virtual screening of ultra-large libraries, predicting toxicity and ADME properties, drug repurposing, optimizing clinical trial designs, and analyzing biological data, with varying degrees of success. Reinforcement learning (RL), a subtype of ML based on rewarding desired behaviors and punishing undesired ones, has been successfully used to generate novel compounds with desirable pharmacological properties in silico (Popova et al., 2018; Wang et al., 2022). Recently, we were able to generate new chemical entities against both native and mutant NAs of influenza A virus (IAV) using a deep RL-based molecule design platform developed by our team (Matevosyan et al., 2022). The RL-generated NCEs had non-sialic acid-like scaffolds and showed no resemblance to existing drugs like oseltamivir. Interestingly, computational experiments revealed that some of these NCEs exhibited excellent binding affinity not only to NAs of IAV but also to the NA of influenza B virus (IBV). These computational results prompted us to further explore the potential applications of deep RL in the design of molecules targeting NAs of both IAV and IBV viruses.

In this study, we extend the application of the same RL-based molecule design platform to design a novel set of compounds with the capability to inhibit NAs of both IAV and IBV. We synthesized nine RL-generated compounds and studied their inhibitory activity in the enzymatic assay. Two compounds emerged as inhibitors of both IAV and IBV NAs at low micromolar concentrations. Furthermore, these compounds demonstrated significant antiviral activity in both *in vitro* and *in vivo*. Our results underscore the efficiency of deep RL in antiviral drug discovery, offering novel avenues for the discovery of effective antiviral agents.

## Materials and Methods

### *RL-based design of molecules*

The main idea behind RL is an agent that acts in an environment and whose goal is to learn the optimal policy in order to maximize a scalar reward signal. In this work, the RL framework is formulated as a Markov Decision Process (MDP). Starting from the initial time step t=0, considering the initial state $S_t$ (in this case, a benzene ring), the agent chooses an action $A_t$ from the set of available actions $A(S_t)$ and receives a reward Rt from the environment. The set of available actions $A(S_t)$ consists of atom- and fragment-based actions. Based on the selected action $A_t$ the molecule corresponding to the current state $S_t$ is modified and state St transitions to a new state $S_{t+1}$. This iterative process continues until either the agent reaches the terminal step at t=T or decides to terminate by choosing the ‘no change’ action. The above-mentioned RL workflow does not rely on any external datasets and instead the agent learns directly from interacting with the environment. This approach enables the unbiased generation of novel molecular scaffolds without relying on existing data. Further information about the MDP structure, implementation details of the agent and the reward shaping are presented in the supplementary materials.

*Chemical similarity calculations*

The chemical similarity calculations of top 31 compounds and oseltamivir are based on RDKit’s implementation of Morgan Fingerprint (radius 2, nBits 2048) (https://www.rdkit.org). All possible pair similarities were calculated and similarities between the same molecules were discarded, thus resulting in 31x31-31=930 similarity values.

*Molecular docking and dynamics simulations*

Crystallographic structures of NAs of IAV and IBV were obtained from the Protein Dat Bank (NA of IAV – PDB ID: 6HP0; 7S0I, and NA of IBV – PDB Id: 4CPY, 4CPO). Molecular docking studies of compounds and oseltamivir were implemented with ICM-Pro software (Abagyan and Totrov, 1994). The “docking effort” and “number of generated conformers” was set to 10 and 15, respectively. The parameter known as "docking effort" quantifies the level of simulation comprehensiveness and directly influences the duration of the simulation. In the context of hydrophobic pockets, it is generally recommended to utilize a value of 1. However, when conducting docking studies involving solvent-exposed pockets or pockets that contain metal ions, this parameter could be increased. In our case, the binding pocket of NA contains a calcium ion. The binding site was defined around the co-crystallized ligand. After the identification of the binding pocket the binding box was built with an offset of 4 Å from each side. In the case of NA of IAV, the height, width and length of the box were 26.5 Å, 25 Å, and 18.8 Å, respectively. For the NA of IBV, the parameters were 24 Å, 25.5 Å and 19.3 Å.

The MD simulations were performed by using GROMACS 2023.1 package with ff14SB and GAFF force fields for protein and ligands parameterizations, respectively (Wang et al., 2004; Maier et al., 2015). The 309.75 K temperature selection aimed to replicate the early phases of a viral disease prior to the manifestation of a fever. The SHAKE algorithm was utilized to maintain the integrity of hydrogen bonds, employing a 2 fs integration step (Ryckaert et al., 1977). By keeping these bonds from stretching or squeezing during the simulation, SHAKE ensures that the simulation stays accurate and realistic. Each simulation comprised a 10 ns period for system equilibration and 30 ns of standard MD employing a simulation time step of 2 fs. The MM/PBSA calculations and RMSD analysis of each MD simulation has been performed using gmx_MMPBSA tool (Valdés-Tresanco et al., 2021). Before performing calculations with gmx_MMPBSA, the GROMACS output trajectory with its periodic boundary conditions was removed.

Cells, compounds and viruses

MDCK (Madin-Darby canine kidney) cells were grown at 37 °C in Eagle’s minimum essential medium (Biowest, France) supplemented with 10% fetal bovine serum (Biowest,

France), 2 mM L-glutamine (Lonza, Switzerland), 100 IU/ml penicillin (Reyoung, China) and 100 µg/ml streptomycin (Arterium, Ukraine).

Compounds selected for in vitro experiments were synthesized by WuXi AppTec (China). Synthesis pathways and details are presented in the supplementary materials. Oseltamivir was purchased from WuXi AppTec.

For IAV studies, we used A/St.Petersburg/63/2020 (H1N1 pdm09), A/Vladivostok/2/2009 (H275Y mutant), and A/WSN/33 (H1N1) strains. For IBV studies, the B/Samara/32/2018 (Yamagata lineage) strain was used. All viruses were a kind gift from the Smorodintsev Research Institute of Influenza (St. Petersburg, Russia).

Enzymatic inhibitory assay

Serial diluted compounds (8 doses, 5-fold dilutions, in duplicate wells) were added into 96-well plates, 25 µL per well. Then 25 µL diluted virus were added into the wells. The compounds and virus were incubated at 37°C for 30 min. After the incubation, diluted MUNANA substrate were added to the wells, 50 µL per well, and the plates were incubated at 37°C for 1 hour. Wells containing the virus and substrate without compound was set as the positive control, while wells containing substrate without compound or virus was set as the negative control. The final DMSO concentration is 0.5%. The reaction was terminated by adding 100 µL of NA-Fluor™ Stop Solution to each well. The plates were detected with microplate readers SpectraMax M2e using an excitation wavelength of 355 nm and an emission wavelength of 460 nm. The raw data were used for calculating the inhibition activity using the equation below:

$$Inhibition\% = 100 - (CPD - Average\ BG)/(Average\ VC - Average\ BG) \times 100\%,$$

where CPD refers to signal of test compounds wells, containing compound + virus + substrate; VC refers to average of signals of virus control wells, containing virus + substrate, no compound; BG refers to average of signals of background control wells, substrate only. $IC_{50}$ values of compounds are calculated with the GraphPad Prism software (version 10) using the nonlinear regression model of log(inhibitor) vs. response - Variable slope.

*Cell-based assays*

Dose-dependent antiviral assay was conducted in a 24-well cell culture plate (seeding dose: $10^5$ cells/well) infected with IAV and IBV strains at MOI of 0.01 $TCID_{50}$/cell and treated with selected compounds at concentrations ranging from 0.5 µM to 50 µM. After 1 h post-infection, the cell culture medium (DMEM/F12 (Biowest, France) supplemented with BSA (Sigma-Aldrich, Germany) and TPCK-trypsin (Sigma-Aldrich, Germany) at 2µg/ml) was replaced with a fresh medium containing the same compounds at indicated concentrations. The infection was allowed to proceed for 10 h when the virus completed the first round of replication. Then, the supernatant was collected and titrated by cytopathic effect (CPE) assay on MDCK cells by 10-fold serial dilutions (Karakus et al., 2018). The $EC_{50}$ values were determined using GraphPad Prism software with the inhibitor vs normalized response - variable slope analysis.

For plaque assay, MDCK cells grown in a 12-well cell culture plate (seeding dose: 2.5 × $10^5$ cells/well) were infected with IAV at 100 $TCID_{50}$/well in the presence of selected compounds at indicated concentrations and oseltamivir. After 1 hours post-infection, cells were washed twice with fresh medium and overlaid with 0.75% agarose (Sigma-Aldrich, Germany). Following 48 hours post-infection, cells were fixed with 10% formaldehyde and stained with 0.5% crystal violet.

For cytotoxicity assay, cell viability was assessed using an MTT (3-[4,5-dimethylthiazol-2-yl]-2,5 diphenyl tetrazolium bromide) assay. MDCK cells seeded in 96-well cell culture plates at a density of 2 × 104 cells/well were exposed to various concentrations of the compounds. Following treatment, the cells were incubated at 37 °C in a 5% CO2 for 24 hours. After the incubation period, the cell culture medium was removed, and the cells were washed with cold PBS. Then, the MTT solution (Sigma-Aldrich, Germany) was added, and the cells were incubated at 37 °C for 2 h. The purple formazan product was then extracted using dimethyl sulfoxide (DMSO). The absorbance of the samples was measured at 570 nm using a microplate reader (BioTek Epoch2, USA). The percentage of viable cells was calculated for each concentration using the formula $[(ODT/ODC) \times 100]$, where $OD_T$ represents the absorbance of the treated cells and $OD_C$ represents the absorbance of the control cells. The concentration of the compound that induced 50% cell death ($CC_{50}$) was determined.

*Western blotting*

MDCK cells were seeded in a 24-well cell culture plate ($10^5$ cells/well), and then infected with IAV at multiplicity of infection of one. After 24 hours post-infection, the cells were lysed in a buffer containing 125 mm Tris-HCl, 5% SDS, 25% glycerol, 0.1% bromophenol blue, and 10% β-mercaptoethanol. Samples were separated by SDS-PAGE and transferred onto pieces of PVDF membranes (Immobilon-P; Merck Millipore Ltd.). To detect viral nucleoprotein and actin, the HRP-conjugated anti-IAV nucleoprotein (C43; Abcam, USA) and anti-actin (AC-15; Abcam, USA) antibodies were used.

*Animal experiments*

For compounds safety experiments, 8-week-old female BALB/c mice (*n* = 6/group) were treated with selected compounds at 100 mg/kg (intraperitoneally), oseltamivir 20 mg/kg (orally) and PBS (intraperitoneally) as control twice daily for 5 days. Animals were monitored for weight loss and mortality for 14 days post-treatment.

For infection experiments, 8-week-old female BALB/c mice (*n* = 6/group) were intranasally inoculated with 50 µl of five 50% mouse lethal doses of IAV and IBV. After 2 hours post-infection, mice in therapeutic groups received selected compounds at 20, 50 and 100 mg/kg concentrations (intraperitoneally), and 20 mg/kg oseltamivir (orally). Control and virus control groups received 1X PBS. All animals were treated twice daily for 5 days and monitored for 14 days. The weight loss was calculated for each mouse as a percentage of its weight before virus inoculation. Animals that displayed weight lost more than 30% of their initial weight were euthanized.

At day 5 post-infection, mice (*n* = 3/group) were sacrificed, and their lungs were homogenized in 1 ml of sterile cold PBS. Then, the titration of collected samples was performed by using CPE assay on MDCK cells by 10-fold serial dilutions (Karakus et al., 2018).

All animal experiments were approved by Institutional Animal Care and Use Committee of the Institute of Molecular Biology of NAS RA (approval #: 14062023/1). This ethics committee is working accordance to International guiding principles for biomedical research involving animals (1985), European convention for the protection of vertebrate animals used for experimental and other scientific purposes, Guide for the Care and Use of Laboratory Animals and other applicable regulation.

*Statistical analysis*

Data were analyzed by using GraphPad Prism software (version 10). Animal data were analyzed by Mantel-Cox test, which was recommended in GraphPad Prism. All virological data was analyzed by Student's t-test. In vitro data are expressed as mean ± SD of three independent experiments.

## Results

*Generation and selection of compounds*

We performed three independent experiments, each using identical initial settings but varying random seeds. The crystallographic holo structures, specifically 6HP0 and 4CPY, were chosen as representative models for NA of IAV and IBV, respectively. This selection was based on two key considerations: (i) minimal conformational differences (< 1 Å) between holo and apo structures (fig. S1), and (ii) the general tendency for holo structures to exhibit superior performance over apo structures in the context of computational drug discovery (McGovern and Shoichet, 2003). All three experiments used the minimized crystallographic structures of NAs of IAV and IBV without the co-crystallized ligands. The process was initiated by transforming a benzene ring into more complex chemical structures as the algorithm learned. The deep RL algorithm systematically explored all feasible modifications of the scaffold, accomplishing this without any prior knowledge of the existing molecules. Subsequently, it employed an epsilon-greedy policy to select a modification for reward calculation, which then served as the basis for the next round of alterations by the algorithm. This iterative process was repeated numerous times until the predefined termination criteria were satisfied. In total, 60291 unique chemical structures were generated. Fig. 1A showed that 86.5% of them had better reward scores than oseltamivir (0.69), which served as the reference threshold for comparison. Furthermore, we observed similar trends for each component of the reward analyzed separately (data not shown), indicating that our platform is capable of generating NCEs with superior physicochemical properties compared to the reference compound.

To assess the generated compounds and narrow down the focus to a manageable number of compounds for in vitro studies, we employed several filters, including the synthetic accessibility (SA) score (Ertl and Schuffenhauer, 2009), quantitative estimate of drug-likeness (QED) (Bickerton et al., 2012), binding free energy, and structural alerts. As a result, we identified 22504 compounds that bore no structural alerts and exhibited higher scores in QED, SA, and binding free energies estimated by AutoDock Vina compared to oseltamivir (fig. 1B). Then, we proceeded by selecting the compounds whose chemical synthesis were available by chemical vendors, resulting in a total of 5543 structures. To assess their potential, we calculated the MM/PBSA score for these compounds using a previously established methodology for short molecular dynamics simulations (MD) (Sahakyan, 2021). It enabled us to further narrow down the list of compounds, focusing specifically on those that exhibited a comparable level of binding free energy with NAs as oseltamivir. As shown in fig. 1C, 31 displayed similar or superior MM/PBSA scores against the NAs of IAV and IBV when compared to oseltamivir.

Based on calculations using the Tanimoto coefficient (Bajusz et al., 2015), the average similarity value among the generated 31 compounds was 0.258. Furthermore, the average similarity to oseltamivir was notably low with a value of 0.13. These results underscore a high level of chemical diversity present among the newly generated structures and no similarity to oseltamivir. Detailed information about all 31 NCEs, including their physicochemical properties and MM/PBSA values is summarized in table S1 and S2.

*Inhibition of NAs by selected compounds*

In order to conduct an enzymatic inhibition assay, a subset of nine compounds, namely DS-22-inf-009, DS-inf-013, DS-inf-016, DS-inf-017, DS-inf-018, DS-22-inf-021, DS-inf-026, DS-inf-030, and DS-inf-031 was selected based on their physicochemical properties and our budget constraints. The inhibitory activity of selected compounds against the NA of IAV A/WSN/33 (H1N1) strain was studied using the NA-Fluor™ Influenza Neuraminidase Assay Kit. Among the nine compounds tested, DS-22-inf-009, DS-inf-016, DS-inf-017, and DS-22-inf-021 displayed inhibitory effects at the tested concentrations (fig. 2A). DS-22-inf-009 and DS-22-inf-021 showed the most potent inhibitory effect with $IC_{50}$ values of 24.8 μM and 11.2 μM, respectively. The slightest effect was observed for DS-inf-017 with $IC_{50}$ values of 66.9 μM. Oseltamivir was used as a reference drug in the NA assay (fig. 2A). Furthermore, we studied the inhibitory effects of DS-22-inf-009 and DS-22-inf-021 on the NA activity of IBV, as these compounds were initially designed to target both IAV and IBV NAs. As expected, DS-22-inf-009 and DS-22-inf-021 suppressed the NA activity of IBV (B/Samara/32/2018 Yamagata lineage) in a dose-dependent manner with $IC_{50}$ values of 21.2 μM and 9.1 μM, respectively (fig. 2B). We also tested these compounds against oseltamivir-resistant IAV A/Vladivostok/2/2009 (H1N1 pdm09) carrying the H275Y mutation. Both DS-22-inf-009 and DS-22-inf-021 exhibited similar inhibitory capabilities against the mutant NA compared to native NAs (fig. 2B). The $IC_{50}$ values of DS-22-inf-009 and DS-22-inf-021 were 18.2 μM and 14.7 μM, respectively.

To study the binding mode and stability of both DS-22-inf-009 and DS-22-inf-021 with the NAs of IAV and IBV, we performed molecular dynamics (MD) simulations using the GROMACS 2023.1 package and crystallographic structures 6HP0 and 4CPY (PDB IDs) for IAV and IBV, respectively. After MD simulations for 30000 ps, the binding pose of oseltamivir in complex with the 6HP0 and 4CPY proteins closely resembled the crystallographic references with the RMSD (Root Mean Square Deviation) values of 0.26 Å and 0.65 Å, respectively (fig. S2). This result indicates the accuracy of protein-ligand complex MD simulations in predicting the precise ligand poses. As shown in fig. 3A, the mean RMSD values for DS-22-inf-009 and DS-22-inf-021 within the active site of the 6HP0 and 4CPY proteins were less than 0.2 nm, suggesting that DS-22-inf-009 and DS-22-inf-021 form stable interactions with the NAs during the simulation period. Furthermore, we observed similar stability when DS-22-inf-009 and DS-22-inf-021 interacted with the crystallographic structures 7S0I and 4CPO, which are apo structures of the NAs of IAV and IBV, respectively (fig. S3). In 6HP0, DS-22-inf-009 established hydrogen bonds involving its sulfonyl and carbamoyl groups with Asp151 and Arg368 (fig. 3B and 3C). In contrast, within 4CPY, the hydrogen bonds occurred with Arg149, Arg291, and Arg373. DS-22-inf-021 was capable of forming hydrogen bonds with Arg293 and Arg368 in 6HP0, while in 4CPY, interactions were observed with Arg291 (fig. 3B and 3D).

These specific amino acid residues also partake in interactions with other NA inhibitors (Wang et al., 2012; Zeller et al., 2017; Matevosyan et al., 2022).

*Antiviral activity of DS-22-inf-009 and DS-22-inf-021*

Based on the NA assay data indicating the inhibitory effect of DS-22-inf-009 and DS-22-inf-021, we further examined their antiviral activity against A/St.Petersburg/63/2020 (H1N1 pdm09), A/Vladivostok/2/2009 (oseltamivir resistant H275Y mutant), and IBV B/Samara/32/2018 Yamagata lineage strains using a cell-based assay. For this purpose, MDCK cells were infected with IAV and IBV strains at MOI of 0.01 $TCID_{50}$/cell and treated with DS-22-inf-009 and DS-22-inf-021 compounds at concentrations ranging from 0.5 μM to 50 μM. The infection was allowed to proceed for 10 hours when the virus completed the first round of replication. Then, the supernatant was collected and titrated by cytopathic effect (CPE) assay by 10-fold serial dilutions. DS-22-inf-009 exhibited a significant dose-dependent inhibitory effect on all strains, resulting in more than 95% reduction in virus yield at 10 and 50 μM concentrations (fig. 4A). The $EC_{50}$ values were found to be 0.56 μM, 2.16 μM, and 2.31 μM against A/St.Petersburg/63/2020, A/Vladivostok/2/2009, and B/Samara/32/2018 strains, respectively. In comparison, DS-22-inf-021 displayed more robust antiviral activity than DS-22-inf-009, exhibiting $EC_{50}$ values of 0.4 μM against A/St.Petersburg/63/2020, 0.29 μM against A/Vladivostok/2/2009, and 0.74 μM against B/Samara/32/2018 strains (fig. 4A). Neither of the compounds elicited cytotoxic effects within the examined concentration range ($CC_{50}$ > 50 μM), thus indicating that the observed antiviral effects were not associated with cytotoxicity (fig. 4B).

The inhibitory effects of both compounds were further observed through western blotting. The synthesis of viral nucleoprotein was considerably diminished in IAV-infected (A/WSN/33) MDCK cells exposed to compounds at concentrations of 5 μM or 25 μM (fig. 4C), which coincides with the plaque assay results demonstrating the reduction of plaques in the presence of compounds at 25 μM concentration (fig. 4D). Based on our *in vitro* results, we assume that these compounds potentially hinder viral replication not solely by NA inhibition, but also through alternative mechanism(s) that contribute to their robust antiviral effects. This assumption is substantiated by the fact that both DS-22-inf-009 and DS-22-inf-021 exhibited heightened antiviral activity in comparison to their inhibitory effects on NAs ($EC_{50} < IC_{50}$).

*Efficiency of DS-22-inf-009 and DS-22-inf-021 in mice*

First, we tested whether DS-22-inf-009 and DS-22-inf-021 could cause adverse effects in mice. Mice receiving compounds at 100 mg/kg concentration did not display weight loss (fig. S4) or changes in clinical signs and behavior during the observation period (data not shown), suggesting a safety profile of both compounds at tested concentration. Then, 8-week-old healthy female mice were intranasally infected A/St.Petersburg/63/2020 and B/Samara/32/2018 strains. After 2 hours of infection, the mice were intraperitoneally injected with different doses of DS-22-inf-009 and DS-22-inf-021 once per day for 5 days. A similar treatment was administered with oseltamivir for comparison purposes. During the course of the infection, untreated control mice exhibited a decline in body-weight and succumbed to infection between 5- and 10-days post-infection (fig. S5A and fig. S5B). In contrast, a dose-dependent recovery of body weight loss was evident in the treatment groups. Mice subjected to DS-22-inf-009 and DS-22-inf-021 at a dosage of 100 mg exhibited reduced weight loss and faster recovery compared to those administered with lower doses (fig. S5A and fig. S5B).

Next, we observed a dose-dependent response to survival in treatment groups infected with IAV and IBV. DS-22-inf-009 administration at 20 mg/kg, 50 mg/kg, and 100 mg/kg provided 15%, 50%, and 65% survival rates against A/St.Petersburg/63/2020 strain, respectively (fig. 5A). Animal groups treated with DS-22-inf-021 showed higher survival rates than DS-22-inf-009 treated groups, reaching 85% protection at 100 mg/kg dose (fig. 5A). For the B/Samara/32/2018 strain, it was observed that 65% and 100% of mice treated with DS-22-inf-009 and DS-22-inf-021 at the highest doses, respectively, survived a lethal viral challenge (fig. 5B).

Finally, we measured the viral titer in lung tissues at 5 days post-infection. Our quantifications revealed that both compounds administered at 100 mg/kg dose exhibited a statistically significant reduction ($p < 0.05$) in the IAV titer within the lung tissues (fig. 5C). Furthermore, both compounds effectively hindered the propagation of IBV, similar to oseltamivir. However, it should be noted that oseltamivir demonstrated more potent antiviral activity as it reduced the viral titers at lower concentrations compared to DS-22-inf-009 and DS-22-inf-021.

## Discussion

The need to develop new antiviral drugs against influenza virus is of paramount importance due to the continual emergence of new viral strains and the potential for drug resistance. However, the protracted timeline often required for the development of novel antiviral drugs underscores the pressing need to accelerate this process, especially in the context of responding to possible pandemics. To address this challenge, the use of innovative technologies is required (Xu et al., 2022; von Delft et al., 2023). With the advancement of computing power and algorithms, it is not surprising that a growing number of researchers are employing ML algorithms either independently or in combination with other computational approaches to identify new antiviral agents (Bess et al., 2022; Floresta et al., 2022). For example, Bung et al. employed RL to optimize deep neural network-based generative and predictive models for the de novo design of small molecules targeting the Mpro enzyme of SARS-CoV-2 (Bung et al., 2021). Similarly, Zhang et al. utilized RL to identify novel molecules with the potential to inhibit both the Mpro and PLpro enzymes of SARS-CoV-2 (Zhang et al., 2022). However, it is important to note that none of the RL-designed antiviral molecules have been subjected to *in vitro* or *in vivo* testing, highlighting the necessity for further validation of RL-based molecule design methods.

In this study, we utilized deep RL to design small molecules targeting the NAs of IAV and IBV. In contrast to other ML approaches, deep RL does not necessarily require any prior training and instead learns directly from the interactions with the environment. This approach allows for unbiased generation of novel scaffolds without the need for existing training datasets. In deep RL the agent, represented by a neural network, interacts with the environment by taking actions based on its current state and subsequently receives rewards and a new state as outcomes of its actions. Our deep RL-based molecule design platform generated more than 60000 unique compounds. Following the filtration process, we synthesized nine compounds and assessed their inhibitory effect in an enzymatic assay. Among these nine compounds, four exhibited dose-dependent inhibition, demonstrating the capability of our RL-based molecule design platform to generate compounds with defined inhibitory properties at a relatively high success rate. The most potent two compounds, DS-22-inf-009 and DS-22-inf-021, were further studied to reveal their antiviral potential in vitro and in vivo. Both compounds inhibited IAV and IBV in a dose-dependent manner with EC50 < 2.5 μM. Based on our in vitro results, we assume that DS-22-inf-009 and DS-22-inf-021 potentially hinder viral replication not solely by NA inhibition, but also through alternative mechanism(s) that contribute to their robust antiviral effects. This assumption is substantiated by the fact that both compounds exhibited heightened antiviral activity in comparison to their inhibitory effects on NAs ($EC_{50} < IC_{50}$). Further studies will unveil all spectrum of antiviral targets engaged by these compounds.

*In vivo* experiments showed that RL-designed compounds protected mice with survival rates ranging from 65% to 100% at a dose of 100 mg/kg, and no signs of toxicity were observed. Despite the administered dose being five times greater than that of oseltamivir, our in vivo results are promising because these compounds were designed using only one iteration, and subsequent optimization may significantly increase their therapeutic activity. Furthermore, their unique chemical structures, distinct from sialic acid, demonstrate the capability of our RL-based molecule design platform to explore novel chemical scaffolds. This is of great significance, as the majority of current efforts in the search for new NA inhibitors predominantly focus on sialic acid containing compounds (Zeller et al., 2017), thus constraining the scope of potential chemical exploration. The discovery of novel scaffolds may have the potential to lead to the development of anti-influenza drugs with an increased resistance barrier in comparison to existing NA inhibitors.

In conclusion, our findings provide evidence supporting the successful implementation of deep RL in the design of novel compounds with antiviral activity. The use of deep RL, either alone or in combination with other computational approaches, holds significant promise for advancing the antiviral drug discovery.

## Acknowledgments

Authors thank Dr. Ruben Abagyan (University of California San Diego, USA) for providing the opportunity to use ICM-Pro software.

## Funding

This work was partially supported by a grant from the Committee of Sciences of the Ministry of Education, Science, Culture and Sports RA (grant #: 23EDP-1F004).

## Conflict of interest statement

The authors have no competing interests to declare.

## Figure Legends

**Figure 1.** Computational evaluation of RL-generated compounds. (A) Distribution of compounds by reward scores. The red line represents the score of oseltamivir. (B) Schematic illustration of the filtration workflow with defined filters and threshold values. (C) The MM/PBSA scores of compounds and oseltamivir with the NAs of IAV (PDB: 6HP0) and IBV (PDB: 4CPY). The blue dashed line indicates the threshold for the selection of NCEs and represents the MM/PBSA score of oseltamivir (-12.56 kcal/mol) with the NAs of IAV minus 10%. (D) Average similarity between the pairs of 31 compounds. The red line represents average similarity to oseltamivir.

**Figure 2.** Interaction of selected compounds with NAs of IAV and IBV. (A) Inhibitory effect of NCEs and oseltamivir on the enzymatic activity of NA. (B) Inhibition of the enzymatic activity of NAs of both H275Y mutant IAV and native IBV Yamagata lineage by DS-22-inf-009 and DS-22-inf-021. Data are expressed as mean ± SD.

**Figure 3.** Computational study of the interaction of DS-22-inf-009 and DS-22-inf-021 with the NAs of IAV and IBV. (A) RMSD analysis of MD simulation trajectory for 30000 ps. (B) The 3D structure of NAs active sites in the presence of compounds and oseltamivir. DS-22-inf-009, DS-22-inf-021, and oseltamivir are colored blue, orange, and yellow, respectively. (C) The interaction map of DS-22-inf-009 with amino acid residues in the NAs active sites. (D) The interaction analysis of DS-22-inf-021 with amino acid residues in the NAs active sites.

**Figure 4.** *In vitro* antiviral activity of DS-22-inf-009 and DS-22-inf-021. (A) Inhibition of wild IAV, oseltamivir-resistant IAV, and IBV by compounds at different concentrations. (B) Viability of MDCK cells upon treatment with DS-22-inf-009 and DS-22-inf-021 at different concentrations. (C) Western blot analysis of nucleoprotein in IAV-infected MDCK cells exposed to compounds at 5 and 25 μM. (D) Plaques of IAV in the presence of DS-22-inf-009 and DS-22-inf-021 at 25 μM.concentration. Results represent the mean (± SD) of three experiments ($n$ = 3). Significant differences compared to control are denoted by *$p < 0.05$ and **$p < 0.01$.

**Figure 5.** Effect of DS-22-inf-009 and DS-22-inf-021 on survival rate and viral load in mice infected with IAV and IBV. Survival rates of mice infected with IAV (A) and IBV (B) treated with compounds and oseltamivir at different doses once daily for 5 days. Survival rates were monitored daily to 14 days post-infection. The survival curves were analyzed by Mantel-Cox test. Curves are significantly different ($P < 0.01$). (C) Viral titers of IAV and IBV detected in lung tissues at 5 days post-infection. Results represent the mean ± SD. Significant differences compared to control are denoted by *$P < 0.05$.

**Figure 1**

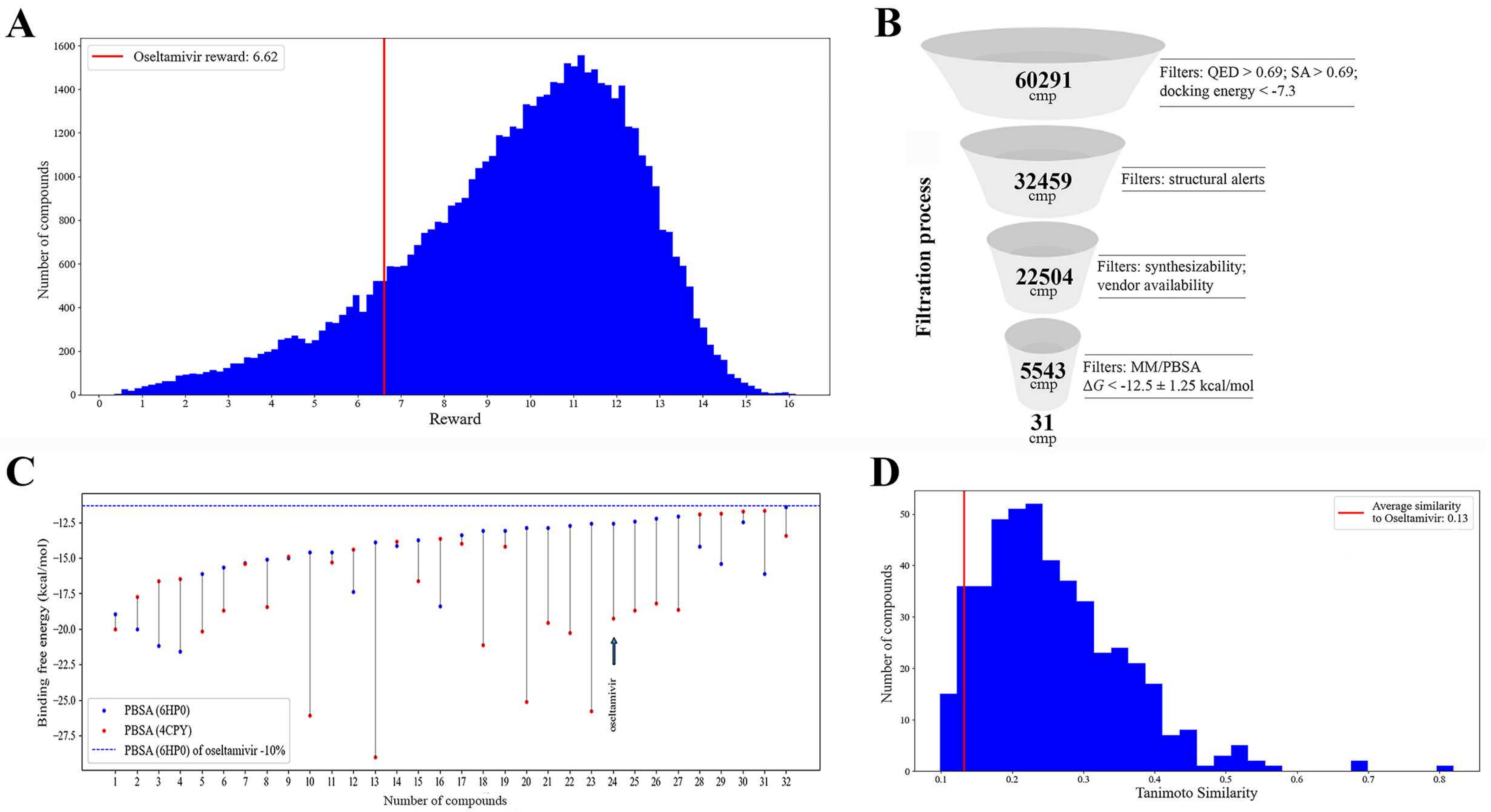

A
Oseltamivir reward: 6.62
Number of compounds
Reward
B
60291 cmp
Filters: QED > 0.69; SA > 0.69; docking energy < -7.3
32459 cmp
Filters: structural alerts
22504 cmp
Filters: synthesizability; vendor availability
5543 cmp
Filters: MM/PBSA ΔG < -12.5 ± 1.25 kcal/mol
31 cmp
Filtration process
C
Binding free energy (kcal/mol)
PBSA (6HP0)
PBSA (4CPY)
PBSA (6HP0) of oseltamivir -10%
oseltamivir
Number of compounds
D
Average similarity to Oseltamivir: 0.13
Number of compounds
Tanimoto Similarity

**Figure 2**

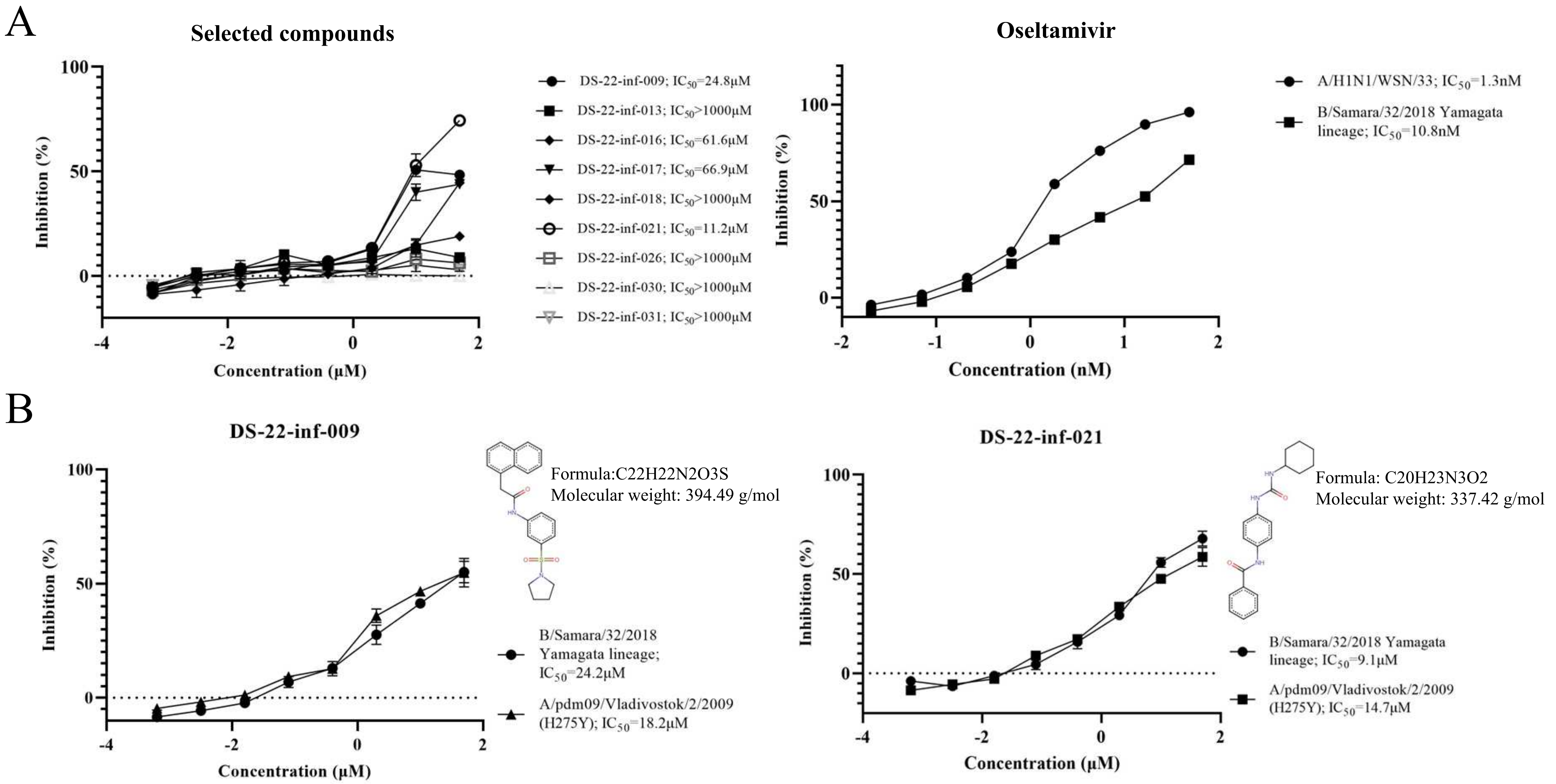
A
Selected compounds
Inhibition (%)
100
50
0
-4
-2
0
2
Concentration (μM)
DS-22-inf-009; IC50=24.8μM
DS-22-inf-013; IC50>1000μM
DS-22-inf-016; IC50=61.6μM
DS-22-inf-017; IC50=66.9μM
DS-22-inf-018; IC50>1000μM
DS-22-inf-021; IC50=11.2μM
DS-22-inf-026; IC50>1000μM
DS-22-inf-030; IC50>1000μM
DS-22-inf-031; IC50>1000μM
Oseltamivir
Inhibition (%)
100
50
0
-2
-1
0
1
2
Concentration (nM)
A/H1N1/WSN/33; IC50=1.3nM
B/Samara/32/2018 Yamagata lineage; IC50=10.8nM
B
DS-22-inf-009
Formula:C22H22N2O3S
Molecular weight: 394.49 g/mol
Inhibition (%)
Concentration (μM)
B/Samara/32/2018 Yamagata lineage; IC50=24.2μM
A/pdm09/Vladivostok/2/2009 (H275Y); IC50=18.2μM
DS-22-inf-021
Formula: C20H23N3O2
Molecular weight: 337.42 g/mol
Inhibition (%)
Concentration (μM)
B/Samara/32/2018 Yamagata lineage; IC50=9.1μM
A/pdm09/Vladivostok/2/2009 (H275Y); IC50=14.7μM

**Figure 3**

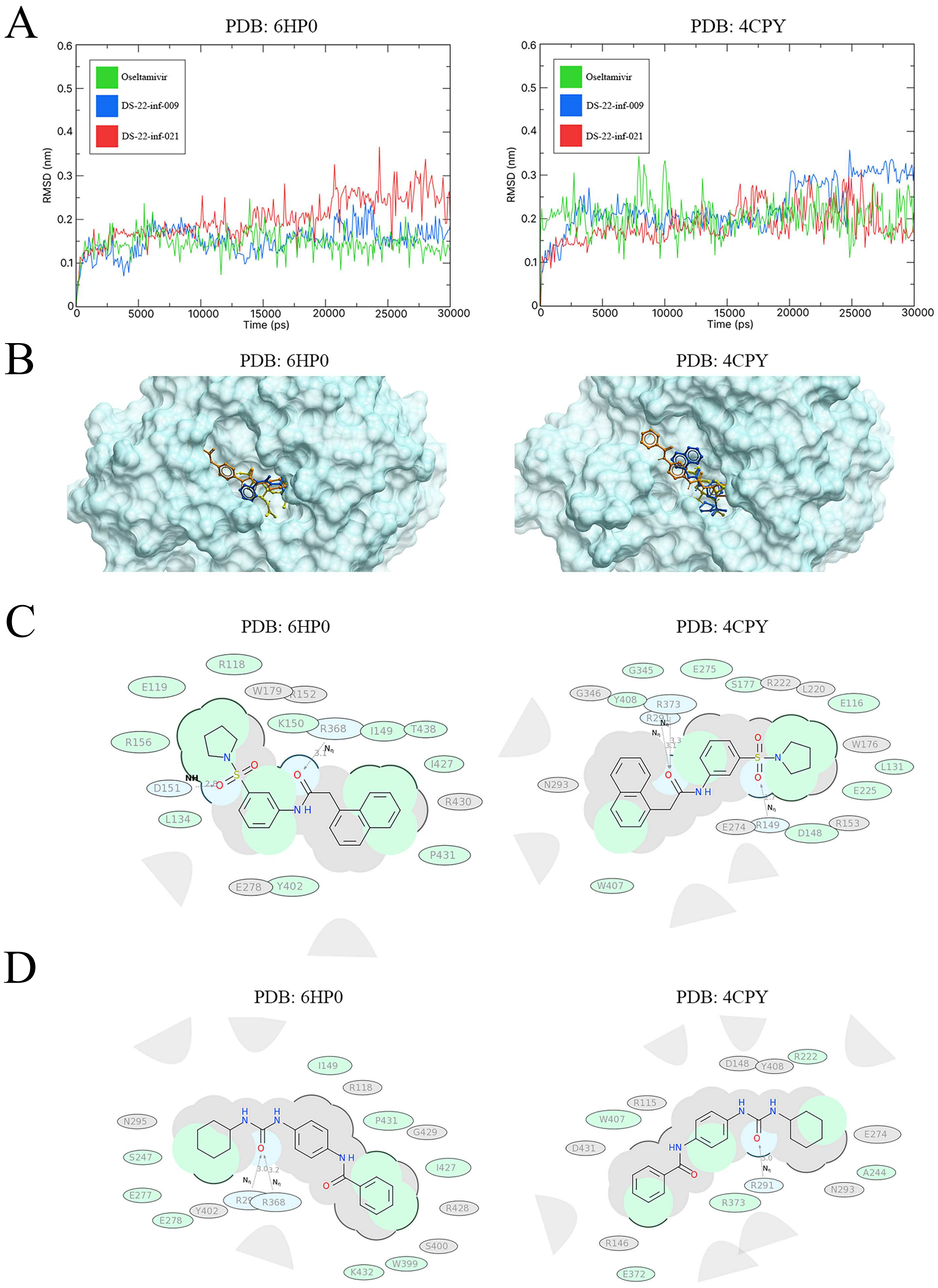
A
PDB: 6HP0
PDB: 4CPY
Oseltamivir
DS-22-inf-009
DS-22-inf-021
RMSD (nm)
Time (ps)
B
PDB: 6HP0
PDB: 4CPY
C
PDB: 6HP0
PDB: 4CPY
D
PDB: 6HP0
PDB: 4CPY

**Figure 4**

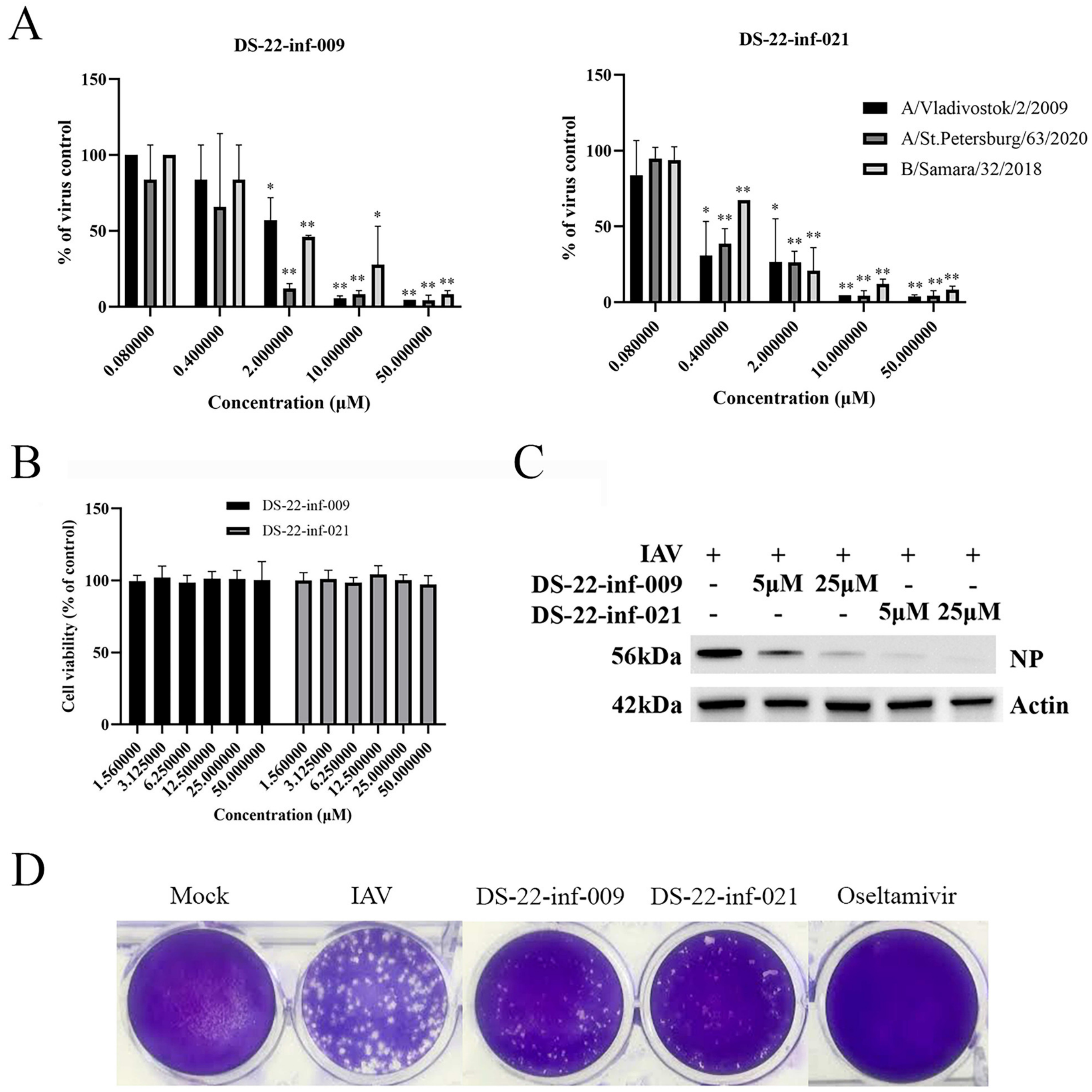
A
DS-22-inf-009
% of virus control
150
100
50
0
0.080000
0.400000
2.000000
10.000000
50.000000
Concentration (μM)
DS-22-inf-021
A/Vladivostok/2/2009
A/St.Petersburg/63/2020
B/Samara/32/2018
B
DS-22-inf-009
DS-22-inf-021
Cell viability (% of control)
1.560000
3.125000
6.250000
12.500000
25.000000
50.000000
Concentration (μM)
C
IAV + + + + +
DS-22-inf-009 - 5μM 25μM - -
DS-22-inf-021 - - - 5μM 25μM
56kDa
NP
42kDa
Actin
D
Mock
IAV
DS-22-inf-009
DS-22-inf-021
Oseltamivir

# Figure 5

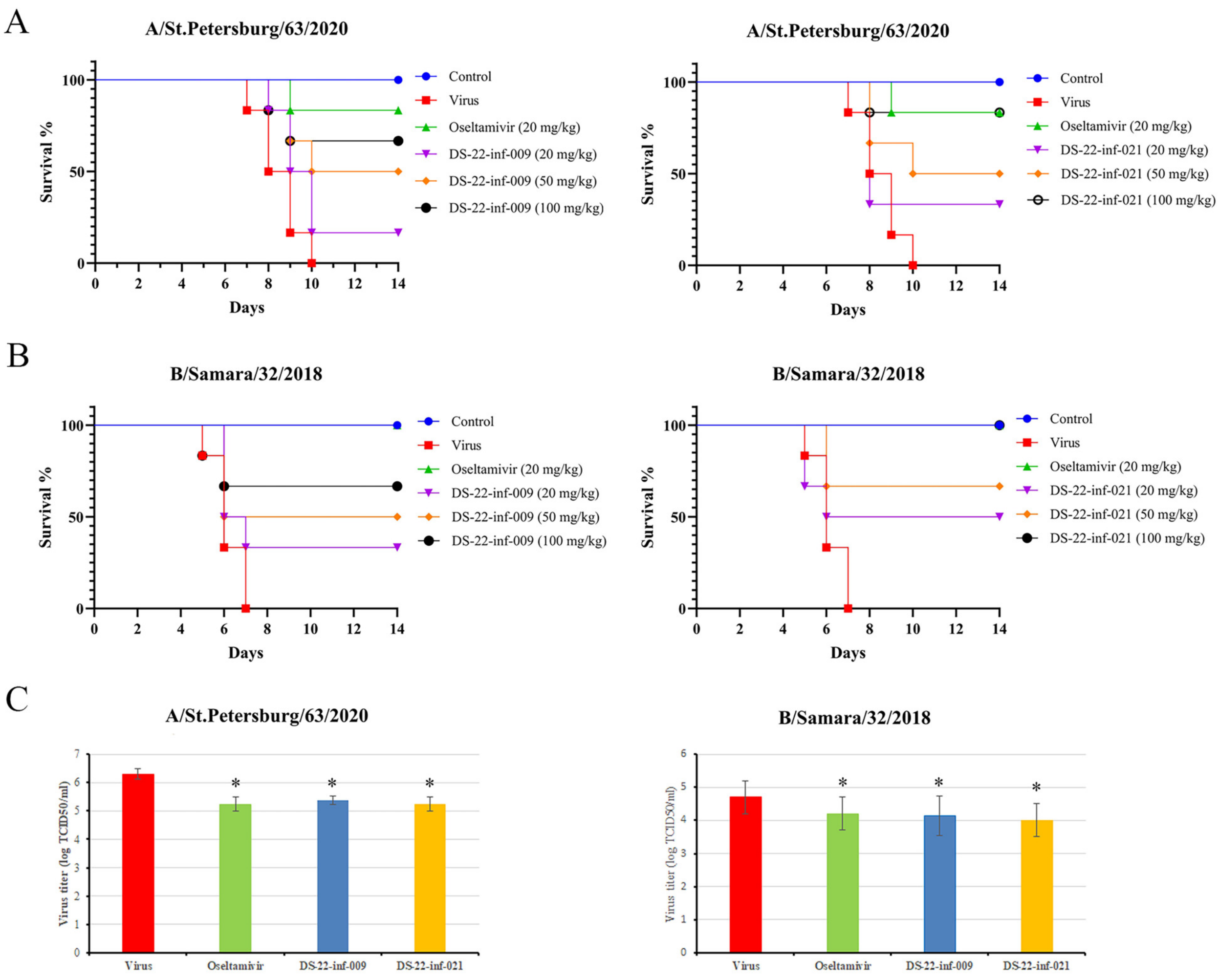
A
A/St.Petersburg/63/2020
Survival %
Days
Control
Virus
Oseltamivir (20 mg/kg)
DS-22-inf-009 (20 mg/kg)
DS-22-inf-009 (50 mg/kg)
DS-22-inf-009 (100 mg/kg)
A/St.Petersburg/63/2020
Survival %
Days
Control
Virus
Oseltamivir (20 mg/kg)
DS-22-inf-021 (20 mg/kg)
DS-22-inf-021 (50 mg/kg)
DS-22-inf-021 (100 mg/kg)
B
B/Samara/32/2018
Survival %
Days
Control
Virus
Oseltamivir (20 mg/kg)
DS-22-inf-009 (20 mg/kg)
DS-22-inf-009 (50 mg/kg)
DS-22-inf-009 (100 mg/kg)
B/Samara/32/2018
Survival %
Days
Control
Virus
Oseltamivir (20 mg/kg)
DS-22-inf-021 (20 mg/kg)
DS-22-inf-021 (50 mg/kg)
DS-22-inf-021 (100 mg/kg)
C
A/St.Petersburg/63/2020
Virus titer (log TCID50/ml)
Virus
Oseltamivir
DS-22-inf-009
DS-22-inf-021
B/Samara/32/2018
Virus titer (log TCID50/ml)
Virus
Oseltamivir
DS-22-inf-009
DS-22-inf-021

**Discovery of new antiviral agents through artificial intelligence: *in vitro* and *in vivo* results**

Roza Izmailyan[1], Mher Matevosyan[2], Hamlet Khachatryan[1,2], Anastasiya Shavina[1,2], Smbat Gevorgyan[1,2], Artur Ghazaryan[1], Irina Tirosyan[2], Yeva Gabrielyan[2], Marusya Ayvazyan[2], Boris Martirosyan[2], Vardan Harutyunyan[2], Hovakim Zakaryan[1,2,*]

[1]Laboratory of Antiviral Drug Discovery, Institute of Molecular Biology of NAS, Hasratyan 7, 0014, Yerevan, Armenia.
[2]Denovo Sciences Inc., 0060, Yerevan, Armenia.

*Corresponding author: Hovakim Zakaryan, e-mail: h_zakaryan@mb.sci.am

## Supplementary Materials

## Materials and methods

### *RL-based design of molecules*

In this work, the RL framework is formulated as a Markov Decision Process (MDP) with the following components:

- State ($S_t$): an undirected graph representing the current molecule, where t is the current time step, indicating the number of modifications the agent has applied starting from the original molecule.
- Action space $A(S_t)$: list of possible modifications that can be applied to the current state. Action At represents a specific modification chosen by the agent. By applying action $A_t$, the current state $S_t$ will transition to a new state $S_{t+1}$, resulting in a modification of the current molecular graph. The action space $A(S_t)$ for the current state $S_t$ or the set of possible modifications is defined as a combination of atom-based similar to MolDQN and fragment-based actions (Zhou et al., 2019). Atom-based actions make small modifications to the current molecule by changing individual atoms or adding/deleting bonds. Fragment-based actions complement atom-based actions by adding or removing entire fragments within the molecule at each step. The fragment-based actions are produced by combining modifications obtained using two frameworks - CReM and mmpdb (Dalke et al., 2018; Polishchuk, 2020). Integration of both atom- and fragment-based methods into the action space enhances agents' capability to navigate the chemical space more effectively. This allows the agent to first utilize the fragment-based approach to identify a chemical subspace suitable for the given protein and then optimize it further with the atom-based approach.
- Reward $R_t$: numerical metric evaluating the benefit of taking action $A_t$ in state $S_t$. Reward is used to encourage or discourage certain actions, guiding the learning process towards optimal decision making.

Similar to the framework described by Matevosyan et al. (2022), we apply several modifications over the standard Deep Q-Network (DQN) architecture (Mnih et al., 2015). Since the current state molecule is represented as an undirected graph, we use a graph neural network (GNN) architecture, Attentive FP, as our policy network (Xiong et al. 2020). Attentive FP uses graph attention mechanism to capture information on local and intermolecular levels. As DQN suffers from overestimated action values that may lead to poorer policies, we employ Double Q-Learning to reduce the overoptimistic value estimates and decouple action selection from action evaluation (Van Hasselt et al., 2016). We use Polyak averaging for updating the target network (Lillicrap et al., 2019). This approach results in a more stable learning process and enhanced sample efficiency compared to periodically copying policy network weights to the target network. We also utilize several techniques to encourage the exploration of the chemical space. Bootstrapped DQN produces useful uncertainty estimates for the agent and leverages these estimates for efficient and deep exploration (Osband et al., 2016). We use decaying epsilon strategy to control the level of exploration and decrease the probability of selecting actions at random as the policy converges. To encourage the agent to explore more chemical space, we change the starting molecule every 100 episodes. For this purpose, we rank all the generated structures by reward and randomly select a new starting molecule from the top 5% percentile. Following the modifications in Matevosyan et al., we implement a Prioritized Experience Replay buffer (Schaul et al., 2016), a strategy that makes learning from experience replay more efficient.

In this study we also modify the learning objective. The existing RL methods mostly optimize the cumulative return over the whole trajectory or the reward received at the end of the episode. In the context of molecular property optimization and *de novo* drug design, it is more suitable to optimize the maximum reward along the trajectory, as opposed to maximizing the expected cumulative return. As this approach is better aligned with the objective of identifying molecules that exhibit favorable properties, we employ the maximum reward formulation version from Gottipati et al. (2020).

During the generation process, the agent receives information about the starting molecule and the 3D structure of the target protein. The agent is not provided with information about the reference ligand structure or any other active molecules. This aspect enables generation of diverse molecules without being biased by existing active ligands.

### *Reward Shaping*

Reward consists of three primary components:

- Binding Energy (BE): to calculate the BE for the generated molecule, we use Autodock Vina docking software (Trott et al., 2010). In this study, we generate molecules targeting NAs of both IAV and IBV viruses. Thus, the BE score for both IAV ($BE_{IAV}$) and IBV ($BE_{IBV}$) are combined through summation. Since the energy score produced by Autodock Vina has a negative sign and as the RL algorithm has a goal to maximize the reward, we adjust the binding energy by multiplying it with -

1:

$$BE = -1 \times \left( BE_{IAV} + BE_{IBV} \right)$$

• Synthesizability score (SA) (Ertl and Schuffenhauer, 2009): to assess the synthesizability of the generated molecules, we utilize a normalized version of the SA Score:

$$SA_{norm} = 1 - \frac{SA - 1}{9}$$

• Drug-likeness: to evaluate drug-likeness, we utilize QED score (Bickerton et al., 2012), which is a metric based on the physicochemical properties of oral drugs. The reward is defined as:

$$R = SA_{norm} \times QED \times BE$$

Both QED and SA scores fall within the range of 0 to 1, whereas the binding energy is not constrained.

*Chemical synthesis*

All synthesis was conducted in WuXi AppTec. H1 NMR spectrum and LC-MS analysis for the all-final compounds were done via Bruker 400 MHz NMR spectrometer and Agilent 1260 HPLC MSD, correspondingly.

DS-22-inf-009

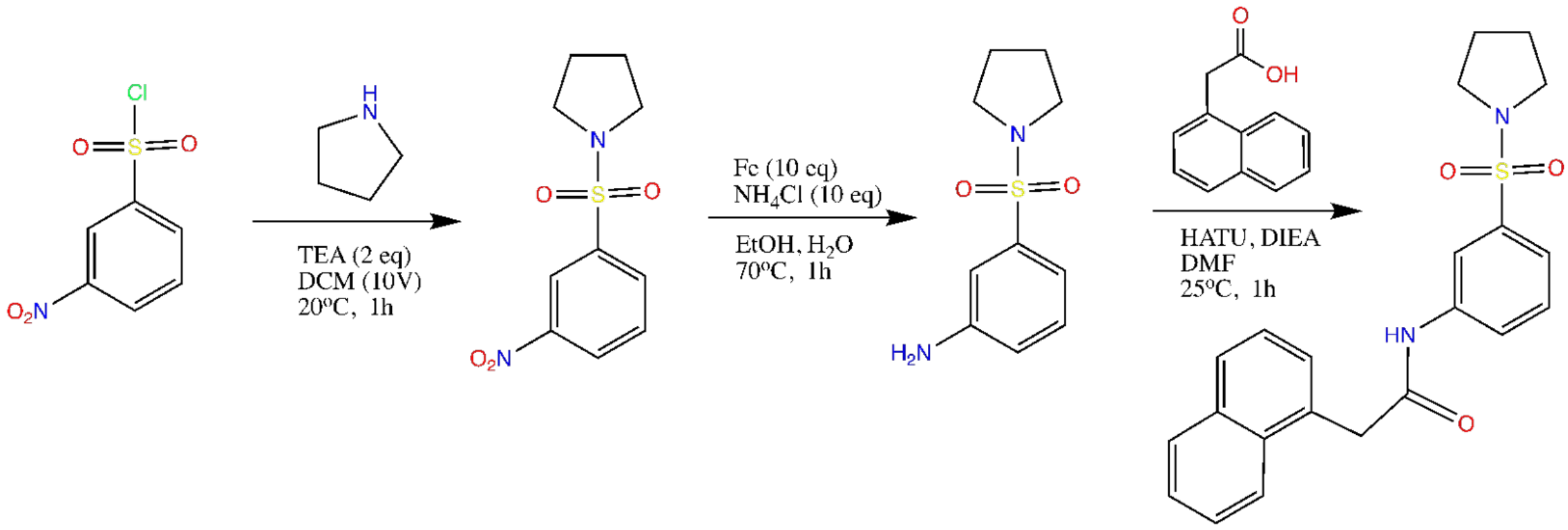

Compound: DS-22-inf-009

Compound DS-22-inf-009 (2-naphthalen-1-yl-N-(3-pyrrolidin-1-ylsulfonylphenyl)acetamide) was synthesized in 3 steps from 3-nitrobenzenesulfonyl chloride via sulfonamide formation, nitro group reduction and further amide bond creation via condensation with 1-naphthylacetic acid according to the scheme mentioned above.

$^1$H NMR (400 MHz, DMSO-*d*) δ 10.70 (s, 1H), 8.16-8.15 (m, 2H), 8.11 – 7.94 (m, 3H), 7.58-7.47 (m, 6H), 4.18 (s, 2H), 3.13 – 3.10 (m, 4H), 1.67 – 1.61 (m, 4H).

DS-22-inf-013

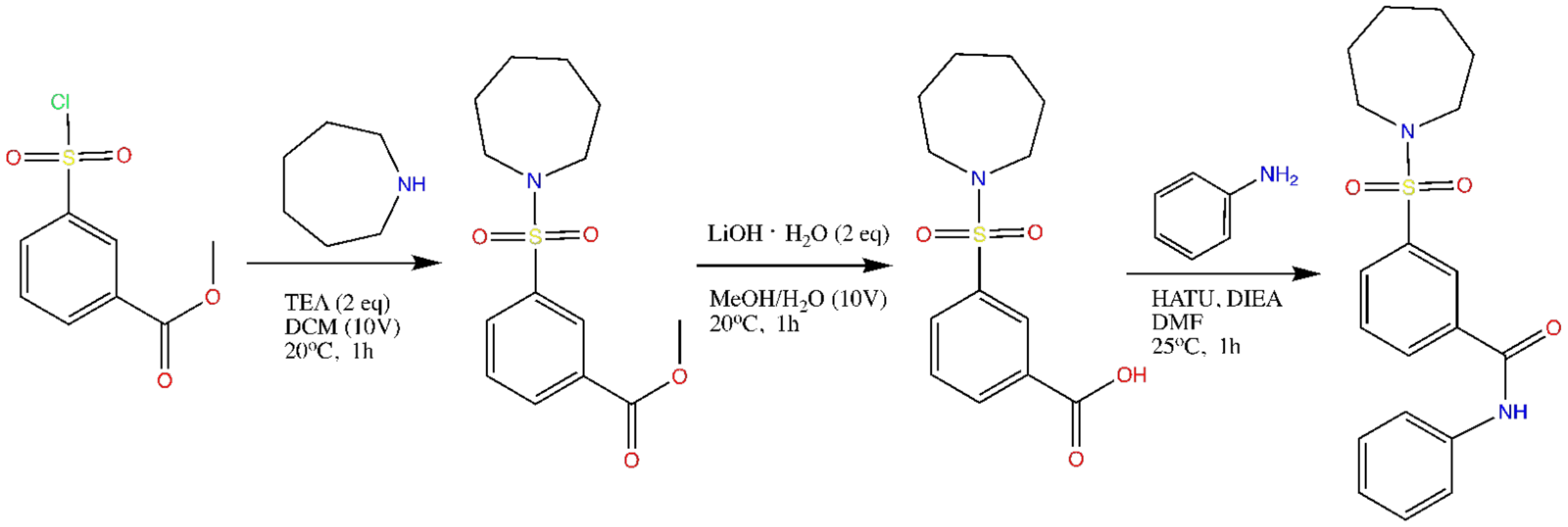

Compound: DS-22-inf-013

Compound DS-22-inf-013 (3-(azepan-1-ylsulfonyl)-N-phenylbenzamide) was synthesized in 3 steps from methyl 3-chlorosulfonylbenzoate via sulfonamide formation with azepane and further
ester hydrolysis in a basic environment. The final step was amide bond formation in reaction with phenylamine according to the scheme mentioned above.

$^{1}$H NMR (400 MHz, DMSO-*d*) δ 10.50 (s, 1H), 8.31 (m, 1H), 8.25-8.23 (m, 1H), 8.00-7.98 (m, 1H), 7.79 – 7.76 (m, 3H), 7.40-7.36 (t, 2H), 7.15 – 7.14 (m, 1H), 3.26-3.23 (t, 4H), 1.64 – 1.52 (m, 4H), 1.51 – 1.50 (m, 4H).

DS-22-inf-016

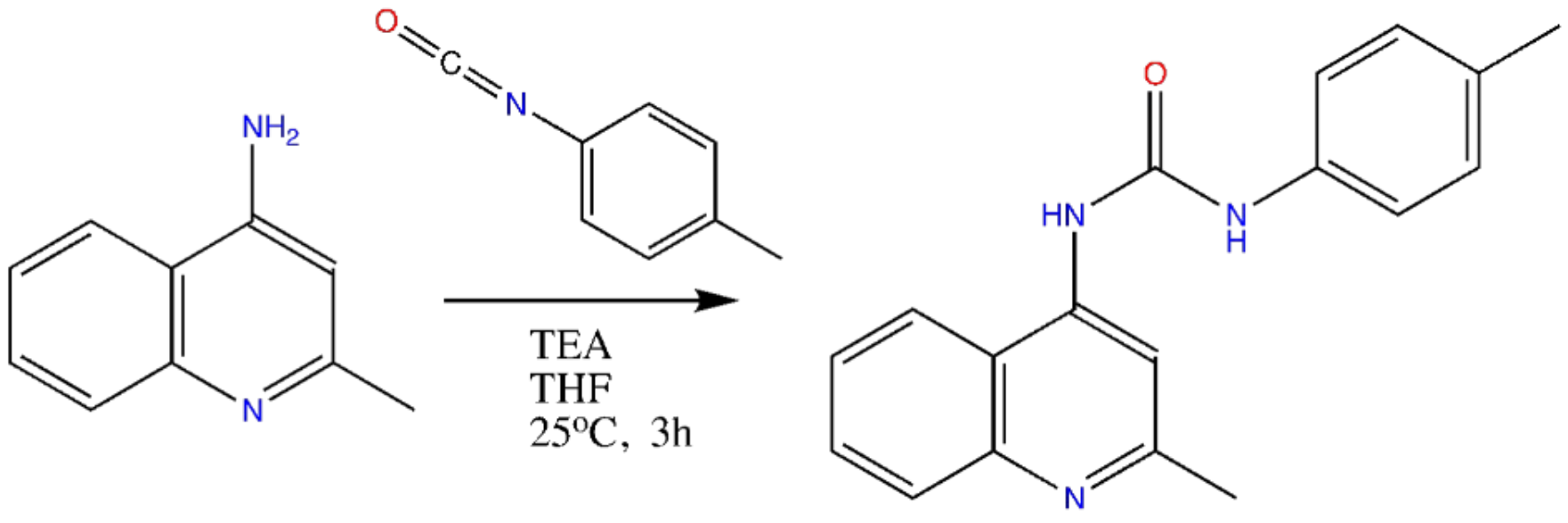

Compound: DS-22-inf-016

Compound DS-22-inf-016 (1-(4-methylphenyl)-3-(2-methylquinolin-4-yl)urea) was synthesized in one step from 2-methylquinolin-4-amine via isocyanate/amine coupling with 1-isocyanato-4-methylbenzene according to the scheme mentioned above.
$^{1}$H NMR (400 MHz, DMSO-*d*) δ 14.98-14.93 (bs, 1H), 10.82 (s, 1H), 10.72 (s, 1H), 8.98 – 8.96 (d, 1H), 8.59 (s, 1H), 8.09 – 8.07 (m, 2H), 7.88-7.84 (t, 1H), 7.48-7.46 (d, 2H), 7.19 – 7.17 (d, 2H), 2.83 (s, 3H), 2.28 (s, 3H).

DS-22-inf-017

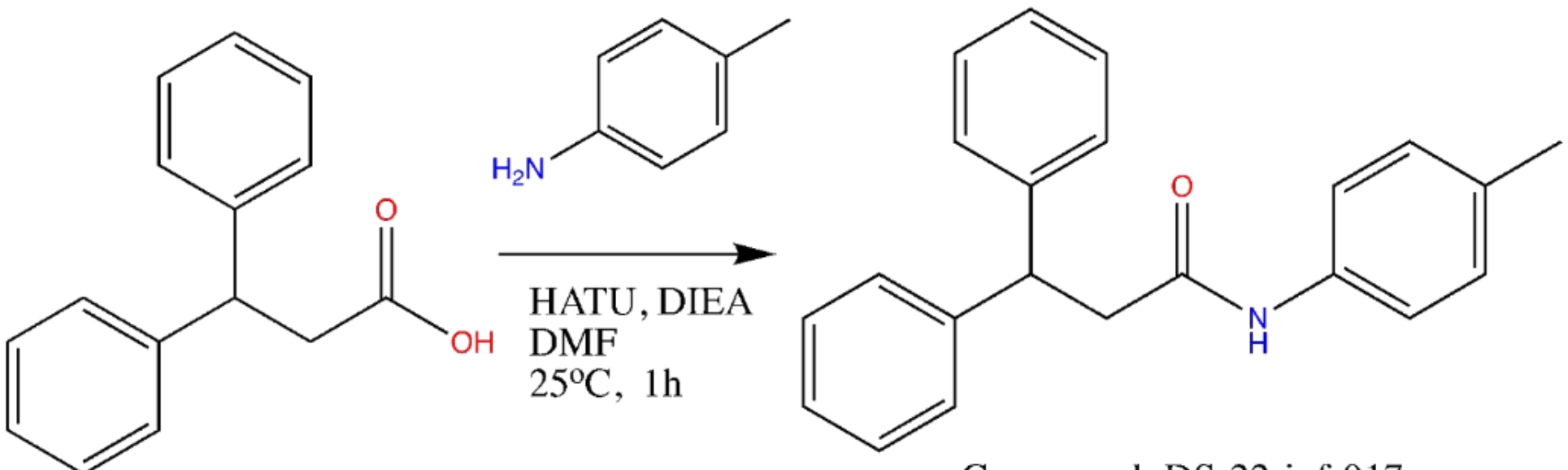

Compound: DS-22-inf-017

Compound DS-22-inf-017 (N-(4-methylphenyl)-3,3-diphenylpropanamide) was synthesized in one step from 3,3-diphenylpropanoic acid via amide formation with 4-methylphenylamine according to the scheme mentioned above.

$^{1}$H NMR (400 MHz, DMSO-*d*) δ 9.86 (s, 1H), 7.37 – 7.27 (m, 10H), 7.15 (m, 2H), 7.05 – 7.03 (m, 2H), 4.59-4.55 (t, 1H), 3.08-3.06 (d, 2H), 2.21 (s, 3H).

DS-22-inf-018

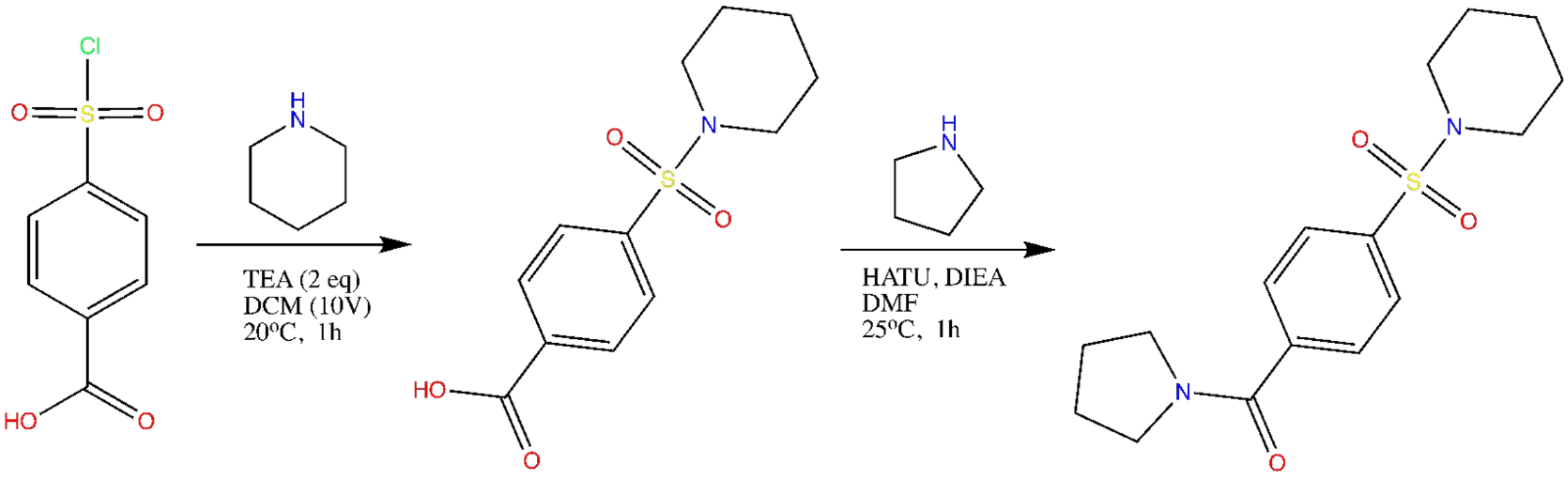

Compound: DS-22-inf-018

Compound DS-22-inf-018 ((4-piperidin-1-ylsulfonylphenyl)-pyrrolidin-1-ylmethanone) was synthesized in steps from 4-chlorosulfonylbenzoic acid via sulfonamide formation with piperidine and further amide formation with pyrrolidine according to the scheme mentioned above.

$^{1}$H NMR (400 MHz, DMSO-*d*) δ 7.78 – 7.71 (m, 4H), 3.48 (t, 2H), 3.34 (t, 2H), 2.93 – 2.89 (m, 4H), 1.90 – 1.79 (m, 4H), 1.55 – 1.50 (m, 4H), 1.39 – 1.36 (m, 2H).

DS-22-inf-021

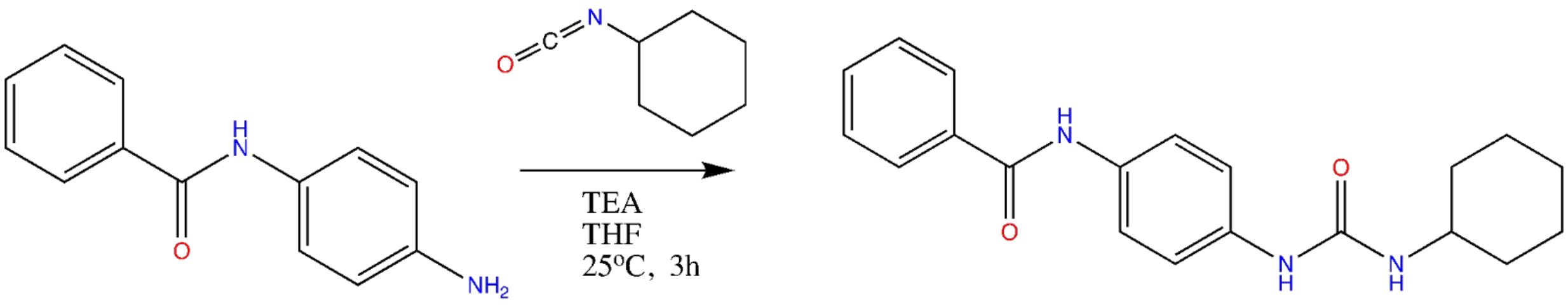

Compound: DS-22-inf-021

Compound DS-22-inf-018 (N-[4-(cyclohexylcarbamoylamino)phenyl]benzamide) was synthesized in one step from N-(4-aminophenyl)benzamide via isocyanate/amine coupling with isocyanatocyclohexane according to the scheme mentioned above.

$^1$H NMR (400 MHz, DMSO-*d*) δ 10.10 (s, 1H), 8.26 (s, 1H), 7.93 (d, 2H), 7.62 – 7.49 (m, 5H), 7.33 (d, 2H), 6.03 (d, 1H), 3.49-3.4 (m, 1H), 1.88-1.76 (m, 2H), 1.74 – 1.61 (m, 2H), 1.59 – 1.44 (m, 1H), 1.38 – 1.19 (m, 2H), 1.18 – 1.05 (m, 3H).

DS-22-inf-026

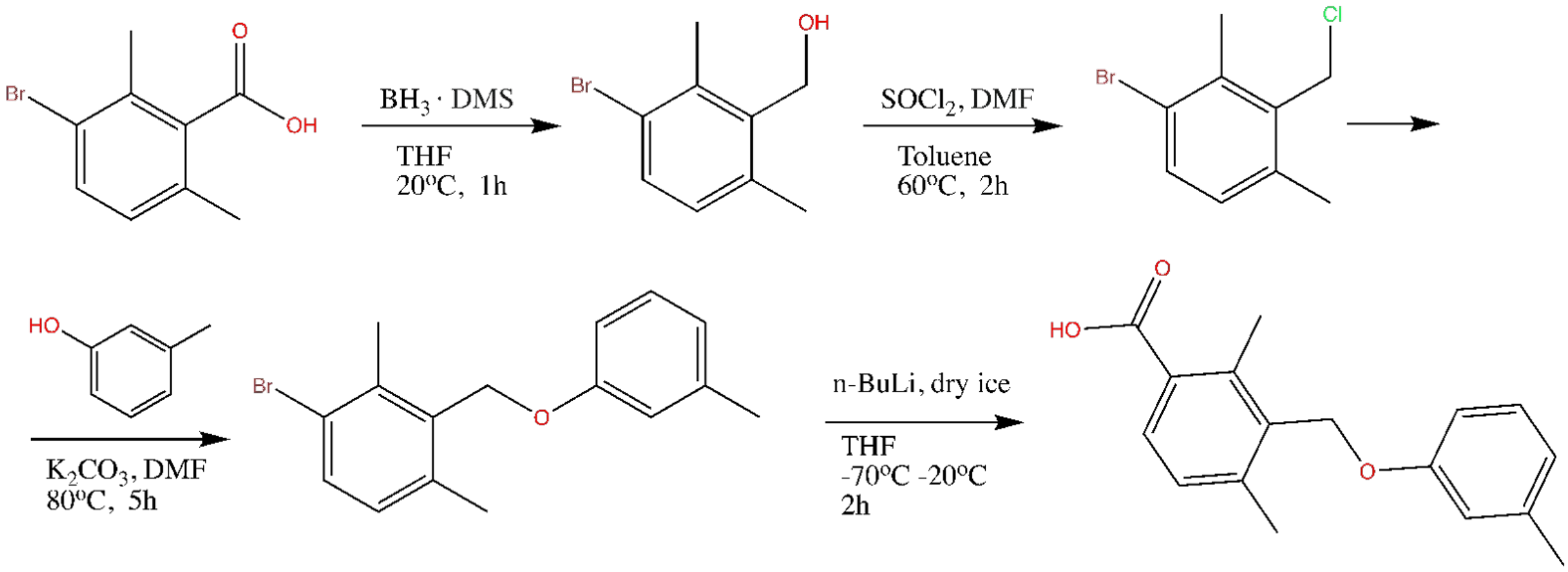

Compound: DS-22-inf-026

Compound DS-22-inf-026 (2,4-dimethyl-3-[(3-methylphenoxy)methyl]benzoic acid) was synthesized in 4 steps from 3-bromo-2,6-dimethylbenzoic acid via acid to alcohol reduction using borane dimethyl sulfide with further alcohol halogenation (using thionyl chloride) and ether formation with 3-methylphenol using Williamson reaction. The final step was carboxylation of bromine according to the scheme mentioned above.

$^1$H NMR (400 MHz, DMSO-*d*) δ 7.35 (d, 1H), 7.18 (t, 1H), 6.98 (d, 1H), 6.90-6.75 (m, 3H), 5.01 (s, 2H), 2.41 (s, 3H), 2.30-2.29 (m, 6H).

DS-22-inf-030

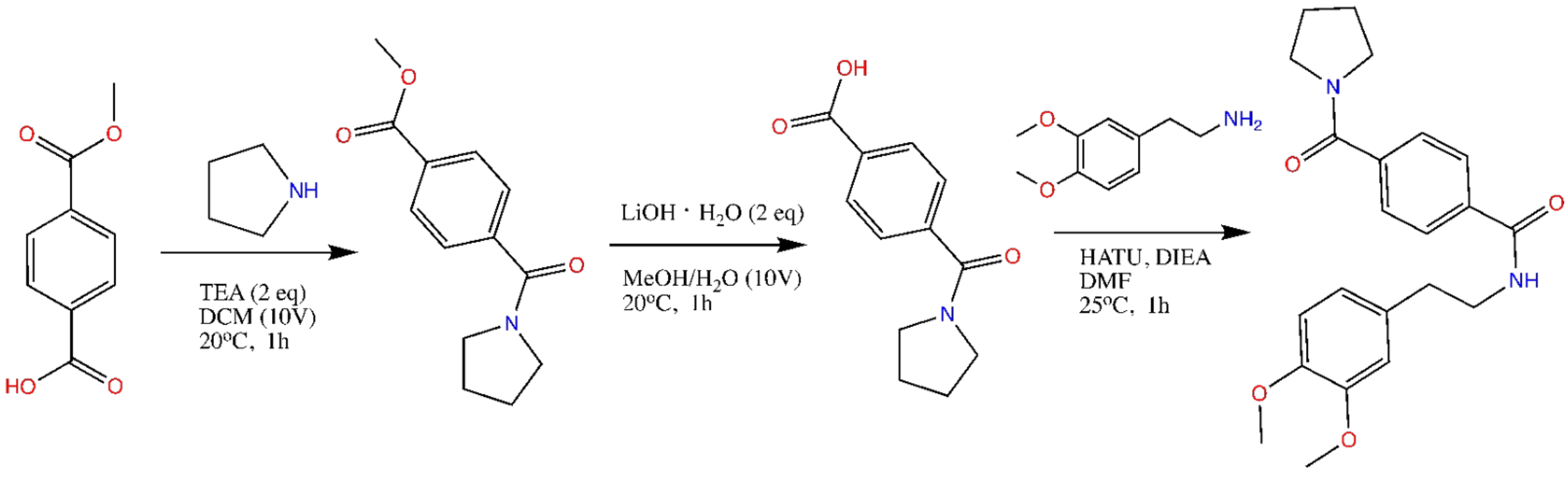

Compound: DS-22-inf-030

Compound DS-22-inf-030 (N-[2-(3,4-dimethoxyphenyl)ethyl]-4-(pyrrolidine-1-carbonyl)benzamide) was synthesized in 3 steps from 4-methoxycarbonyl benzoic acid amide formation with pyrrolidine and further ester hydrolysis in a basic environment. The final step was amide bond formation in reaction with 2-(3,4-dimethoxyphenyl)ethanamine according to the scheme mentioned above.

[1]H NMR (400 MHz, DMSO-*d*) δ 8.61 (t, 1H), 7.85 (d, 2H), 7.57 (d, 2H), 6.87-6.82 (m, 2H), 6.76-6.73 (m, 1H), 3.71 (s, 6H), 3.47 (t, 4H), 3.35 (t, 2H), 2.78 (t, 2H), 1.88 – 1.79 (m, 4H).

DS-22-inf-031

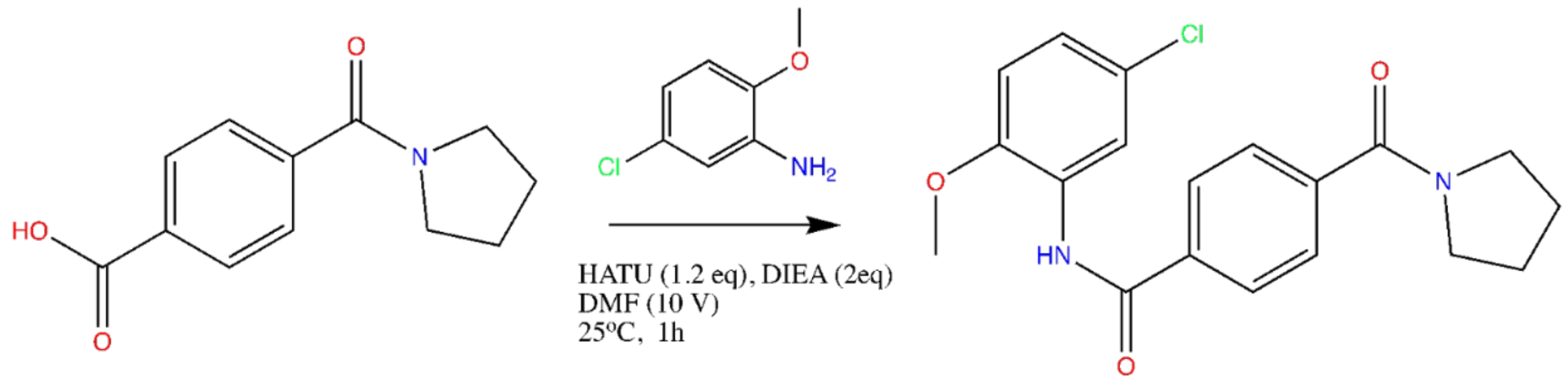

Compound: DS-22-inf-031

Compound DS-22-inf-032 (N-(5-chloro-2-methoxyphenyl)-4-(pyrrolidine-1-carbonyl)benzamide) was synthesized in one step from 4-(pyrrolidine-1-carbonyl)benzoic acid via amide formation with 5-chloro-2-methoxyaniline according to the scheme mentioned above.
[1]H NMR (400 MHz, DMSO-*d*) δ 9.62 (s, 1H), 8.03 – 7.97 (m, 2H), 7.90 – 7.80 (m, 1H), 7.64 (d, 2H), 7.23 (dd, 1H), 7.13 (d, 1H), 3.85 (s, 3H), 3.48 (t, 2H), 3.37 (t, 2H), 1.90 – 1.81 (m, 4H).
Abbreviations: TEA – Triethylamine; DCM – Dichloromethane; HATU - Hexafluorophosphate Azabenzotriazole Tetramethyl Uronium; DIEA - *N,N* Diisopropylethylamine; DMF – Dimethylformamide; THF - Tetrahydrofuran ; DMS - Dimethyl sulfide .

**Fig. S3.** Body weight of mice ($n$ = 6/group) infected with IAV (A) and IBV (B) and treated with compounds at indicated concentrations. Animals received the treatment twice daily for 5 days. Body weights were monitored over 14 days. Data are expressed as mean ± SD.

## SI References

## Figure Legends

**Figure S1.** Apo and holo structures of the NAs of IAV and IBV. Structures (*A* chains) were cleaned from water and other molecules and superimposed based on interacting amino acids via the PyMol 2.51 visualization toolkit (Schrödinger and DeLano, 2020). The RMSDs were calculated between the interacting amino acids via PyMol. Interacting amino acids were identified using the Protein-Ligand Interaction Profiler (PLIP) web service (Adasme et al., 2021).

**Figure S2.** Interaction of oseltamivir with the NAs of IAV and IBV evaluated by MD simulaions. (A) The 3D structure of NAs active sites in the presence of docked (yellow) and reference (green) oseltamivir. (B) The interaction map of oseltamivir with amino acid residues in the NAs active sites.

**Figure S3.** RMSD analysis of MD simulation trajectory for 50000 ps. 7S0I and 4CPO are apo structures of the NAs of IAV and IBV, respectively.

**Figure S4.** Safety profile of compounds in mice. Animals (*n* = 6/group) received compounds at indicated concentrations twice daily for 5 days. Body weights were monitored during 14 days. Data are expressed as mean ± SD.

**Figure S5.** Body weight of mice (*n* = 6/group) infected with IAV (A) and IBV (B) and treated with compounds at indicated concentrations. Animals received the treatment twice daily for 5 days. Body weights were monitored over 14 days. Data are expressed as mean ± SD.

**Table S1.** List of selected compounds, their SMILES and chemical similarity to anti-influenza drugs measured by Tanimoto coefficient (TC). TC varies from 0, when the fingerprints have no commo bits, to 1, when all fingerprints are identical.

| ID/Name | SMILES | Similarity to Oseltamivir | Similarity to Zanamivir | Similarity to Peramivir | Similarity to Laninamivir | Similarity to Baloxavir |
|---|---|---|---|---|---|---|
| DS-22-inf-001 | Cc1ccc(S(=O)(=O)NC(=O)Nc2cccc(Br)c2)cc1 | 0.12 | 0.12 | 0.08 | 0.11 | 0.07 |
| DS-22-inf-002 | Cc1cccc(C(=O)Nc2cccc(CO)c2)c1 | 0.13 | 0.13 | 0.08 | 0.12 | 0.07 |
| DS-22-inf-003 | COc1cccc(NC(=O)C2CCN(C(=O)c3ccccc3)CC2)c1 | 0.14 | 0.11 | 0.10 | 0.13 | 0.12 |
| DS-22-inf-004 | O=C(c1ccccc1)N1CCN(C(=O)c2ccc(O)cc2)CC1 | 0.11 | 0.09 | 0.08 | 0.08 | 0.14 |
| DS-22-inf-005 | O=C(c1ccc(Cl)c(Cl)c1)N1CCN(c2cccc(F)c2)CC1 | 0.09 | 0.06 | 0.06 | 0.06 | 0.13 |
| DS-22-inf-006 | Cc1cc(C)cc(NC(=O)c2ccc(C(=O)N3CCOCC3)cc2)c1 | 0.14 | 0.12 | 0.08 | 0.11 | 0.13 |
| DS-22-inf-007 | COc1ccc(OC)c(NC(=O)C2CCN(C(=O)c3ccccc3)CC2)c1 | 0.14 | 0.10 | 0.10 | 0.13 | 0.11 |
| DS-22-inf-008 | CC(=O)N1CCC(C(=O)Nc2ccc(CC(=O)O)cc2)CC1 | 0.17 | 0.15 | 0.14 | 0.14 | 0.10 |
| DS-22-inf-009 | O=C(Cc1cccc2ccccc12)Nc1cccc(S(=O)(=O)N2CCCC2)c1 | 0.11 | 0.08 | 0.07 | 0.08 | 0.11 |
| DS-22-inf-010 | O=C(Nc1nc2ccccc2[nH]1)c1cccc(Cl)c1 | 0.08 | 0.07 | 0.05 | 0.07 | 0.10 |
| DS-22-inf-011 | Cc1cccc(C(=O)N2CCC(C(=O)NCc3ccc(Cl)cc3)CC2)c1 | 0.15 | 0.10 | 0.09 | 0.10 | 0.13 |
| DS-22-inf-012 | Cc1cccc(C(=O)N2CCC(C(=O)NCc3ccccc3)CC2)c1 | 0.16 | 0.12 | 0.11 | 0.12 | 0.12 |
| DS-22-inf-013 | O=C(Nc1ccccc1)c1cccc(S(=O)(=O)N2CCCCCC2)c1 | 0.10 | 0.09 | 0.08 | 0.08 | 0.11 |
| DS-22-inf-014 | Cc1cccc(C(=O)Nc2sc3c(c2C(N)=O)CCCC3)c1 | 0.14 | 0.12 | 0.11 | 0.12 | 0.12 |
| DS-22-inf-015 | Cc1cccc(OCc2c(C)ccc(C(=O)O)c2C)c1 | 0.15 | 0.12 | 0.08 | 0.13 | 0.07 |
| DS-22-inf-016 | Cc1ccc(NC(=O)Nc2cc(C)nc3ccccc23)cc1 | 0.14 | 0.13 | 0.08 | 0.16 | 0.08 |
| DS-22-inf-017 | Cc1ccc(NC(=O)CC(c2ccccc2)c2ccccc2)cc1 | 0.17 | 0.17 | 0.12 | 0.06 | 0.08 |
| DS-22-inf-018 | O=C(c1ccc(S(=O)(=O)N2CCCCC2)cc1)N1CCCC1 | 0.10 | 0.07 | 0.05 | 0.10 | 0.09 |
| DS-22-inf-019 | O=C(NCc1ccc(C(=O)N2CCCC2)cc1)NCC1CCCCC1 | 0.13 | 0.10 | 0.09 | 0.10 | 0.10 |
| DS-22-inf-020 | O=C(Nc1ccc(C(=O)Nc2ccc3c(c2)OCO3)cc1)c1ccccc1 | 0.10 | 0.11 | 0.08 | 0.12 | 0.10 |
| DS-22-inf-021 | O=C(Nc1ccc(NC(=O)c2ccccc2)cc1)NC1CCCCC1 | 0.13 | 0.12 | 0.10 | 0.12 | 0.10 |
| DS-22-inf-022 | Cc1cc(C)cc(C(=O)Nc2ccc(NC(=O)C3CC3)cc2)c1 | 0.15 | 0.13 | 0.10 | 0.12 | 0.07 |
| DS-22-inf-023 | O=C(c1cccc(S(=O)(=O)Nc2ccccc2)c1)N1CCCCC1 | 0.10 | 0.09 | 0.08 | 0.08 | 0.13 |
| DS-22-inf-024 | O=C(CN1CCN(C(=O)c2ccc(-c3ccccc3)cc2)CC1)N1CCCC1 | 0.11 | 0.09 | 0.08 | 0.09 | 0.11 |
| DS-22-inf-025 | CC(=O)Nc1ccc(NC(=O)c2cccc(O)c2)cc1 | 0.14 | 0.13 | 0.10 | 0.13 | 0.09 |
| DS-22-inf-026 | O=C(Nc1cccc(C(=O)N2CCCC2)c1)C1CCCCC1 | 0.12 | 0.09 | 0.08 | 0.08 | 0.12 |
| DS-22-inf-027 | O=C(Nc1ccc(NC(=O)c2ccc(Cl)cc2)cc1)c1ccccc1 | 0.09 | 0.09 | 0.07 | 0.10 | 0.09 |
| DS-22-inf-028 | Cc1cccc(C(=O)Nc2ccc(NC(=O)C3CCCC3)cc2)c1 | 0.14 | 0.13 | 0.09 | 0.10 | 0.09 |
| DS-22-inf-029 | Cc1ccc(S(=O)(=O)CC(=O)Nc2ccc(C(F)(F)F)cc2)cc1 | 0.15 | 0.09 | 0.09 | 0.13 | 0.06 |
| DS-22-inf-030 | COc1ccc(CCNC(=O)c2ccc(C(=O)N3CCCC3)cc2)cc1OC | 0.14 | 0.10 | 0.10 | 0.13 | 0.10 |
| DS-22-inf-031 | COc1ccc(Cl)cc1NC(=O)c1ccc(C(=O)N2CCCC2)cc1 | 0.13 | 0.10 | 0.07 | 0.11 | 0.09 |

**Table S2.** Computational scores and properties of selected compounds and oseltamivir.

| ID/Name | QED | SA | Reward | MM/PBSA (6HP0) (kcal/mol) | MM/PBSA (4CPY)(kcal/mol) |
|---|---|---|---|---|---|
| DS-22-inf-001 | 0.87 | 0.91 | 11.85 | -12.5729 | -25.793 |
| DS-22-inf-002 | 0.86 | 0.94 | 12.51 | -12.2206 | -18.1967 |
| DS-22-inf-003 | 0.93 | 0.92 | 13.65 | -18.9665 | -20.028 |
| DS-22-inf-004 | 0.92 | 0.93 | 13.22 | -13.0662 | -14.1629 |
| DS-22-inf-005 | 0.81 | 0.91 | 10.89 | -13.0689 | -21.109 |
| DS-22-inf-006 | 0.93 | 0.9 | 13.61 | -21.1796 | -16.6122 |
| DS-22-inf-007 | 0.88 | 0.91 | 12.73 | -21.5648 | -16.4697 |
| DS-22-inf-008 | 0.88 | 0.91 | 12.15 | -18.406 | -13.6352 |
| DS-22-inf-009 | 0.71 | 0.89 | 12.52 | -12.0485 | -18.6439 |
| DS-22-inf-010 | 0.74 | 0.91 | 10.39 | -12.853 | -19.5601 |
| DS-22-inf-011 | 0.88 | 0.91 | 13.74 | -16.119 | -20.1379 |
| DS-22-inf-012 | 0.93 | 0.92 | 14.12 | -15.6322 | -18.6911 |
| DS-22-inf-013 | 0.9 | 0.92 | 13.85 | -14.6067 | -26.08 |
| DS-22-inf-014 | 0.91 | 0.88 | 12.88 | -12.4228 | -18.7022 |
| DS-22-inf-015 | 0.91 | 0.9 | 12.78 | -15.0814 | -18.4378 |
| DS-22-inf-016 | 0.73 | 0.91 | 10.34 | -14.1308 | -13.8242 |
| DS-22-inf-017 | 0.69 | 0.93 | 9.92 | -14.1891 | -11.9011 |
| DS-22-inf-018 | 0.85 | 0.91 | 11.61 | -13.8736 | -29.0356 |

**Figure S1**

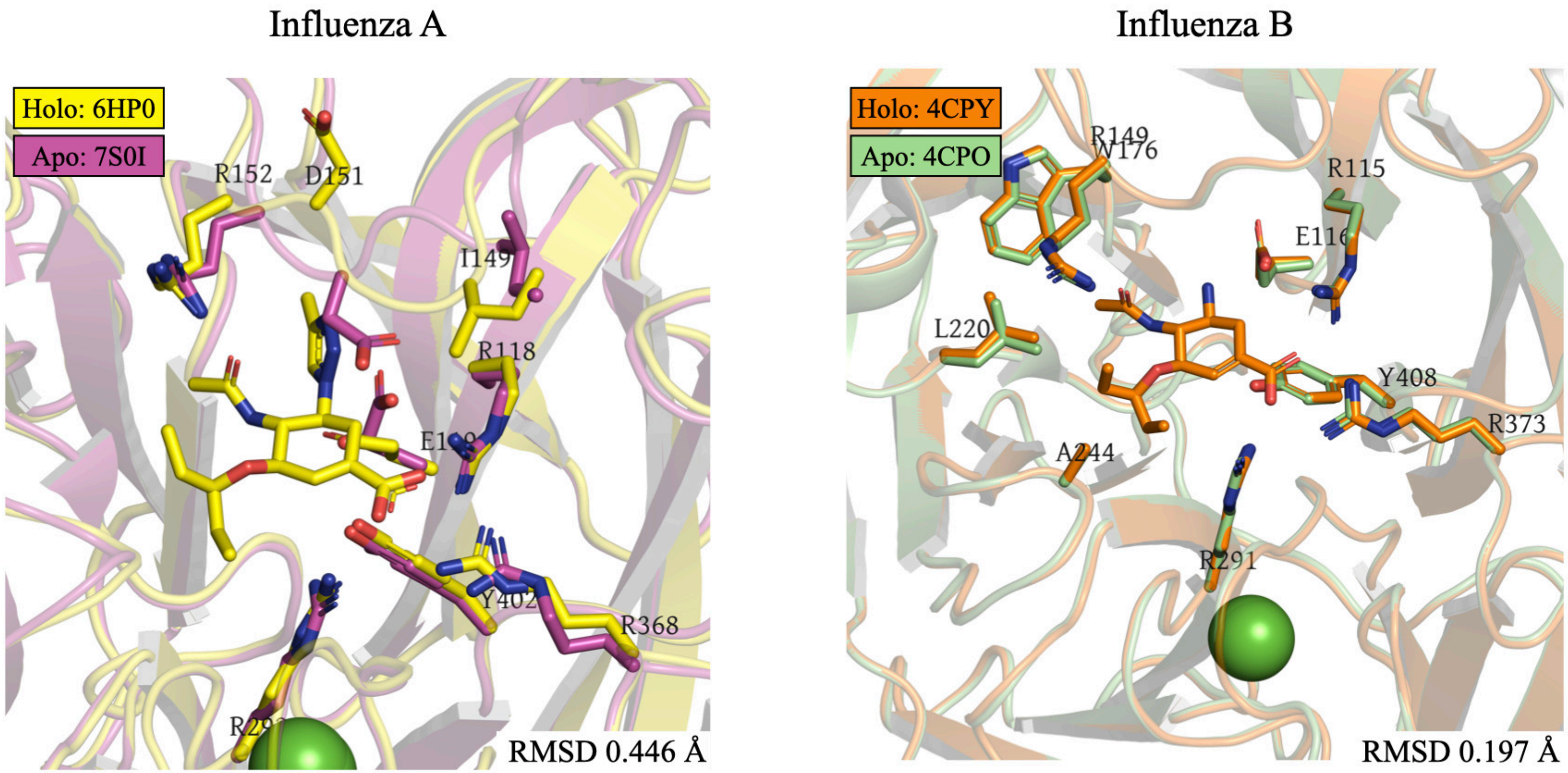
Influenza A
Holo: 6HP0
Apo: 7S0I
R152
D151
I149
R118
E119
Y402
R368
RMSD 0.446 Å
Influenza B
Holo: 4CPY
Apo: 4CPO
R149
W176
R115
E116
L220
Y408
R373
A244
R291
RMSD 0.197 Å

**Figure S2**

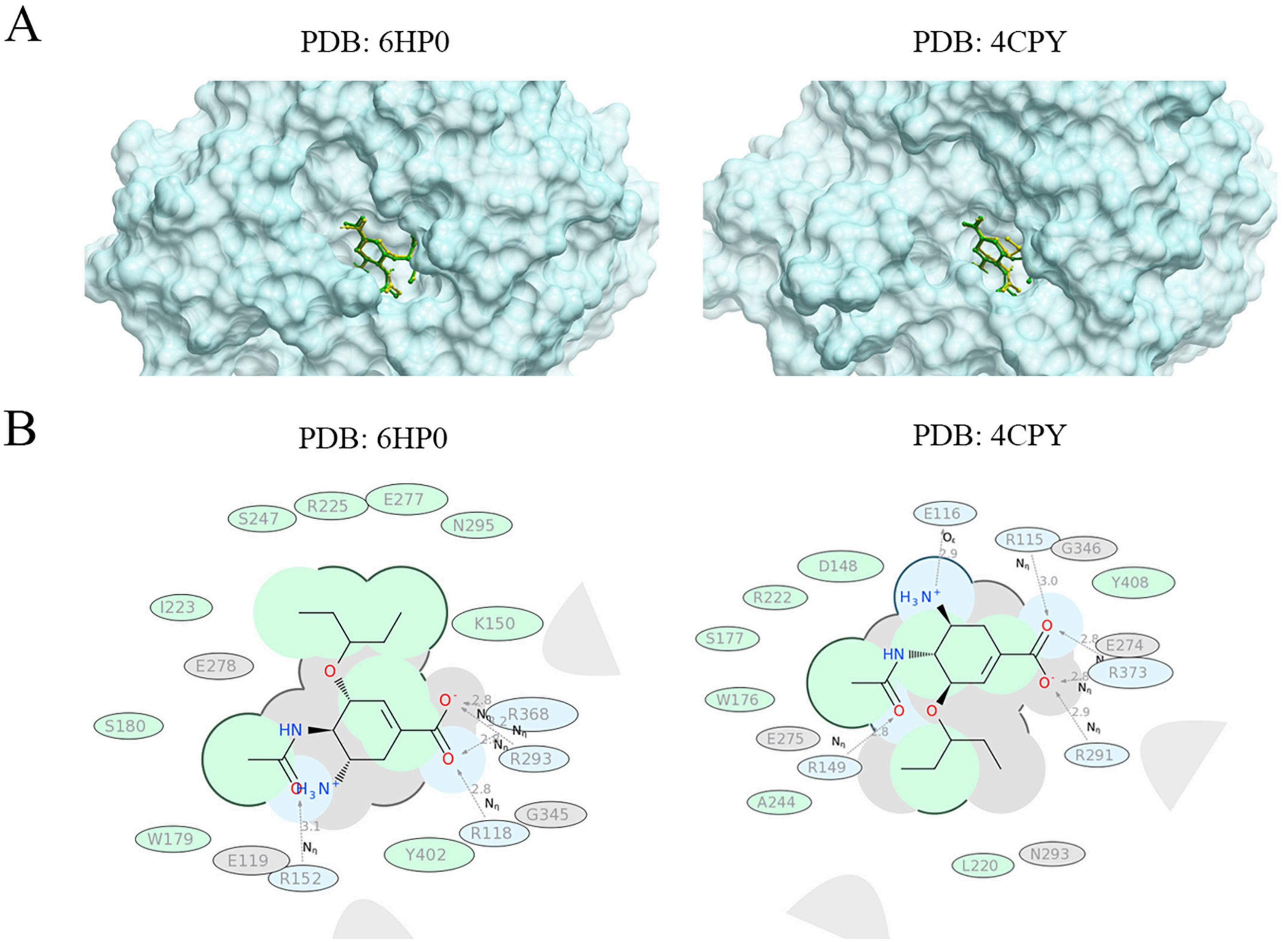
A
PDB: 6HP0
PDB: 4CPY
B
PDB: 6HP0
PDB: 4CPY
S247
R225
E277
N295
I223
K150
E278
S180
R368
R293
G345
R118
Y402
W179
E119
R152
HN
O
H3N+
E116
R115
G346
D148
Y408
R222
S177
E274
R373
W176
E275
R149
R291
A244
L220
N293

## Figure S3

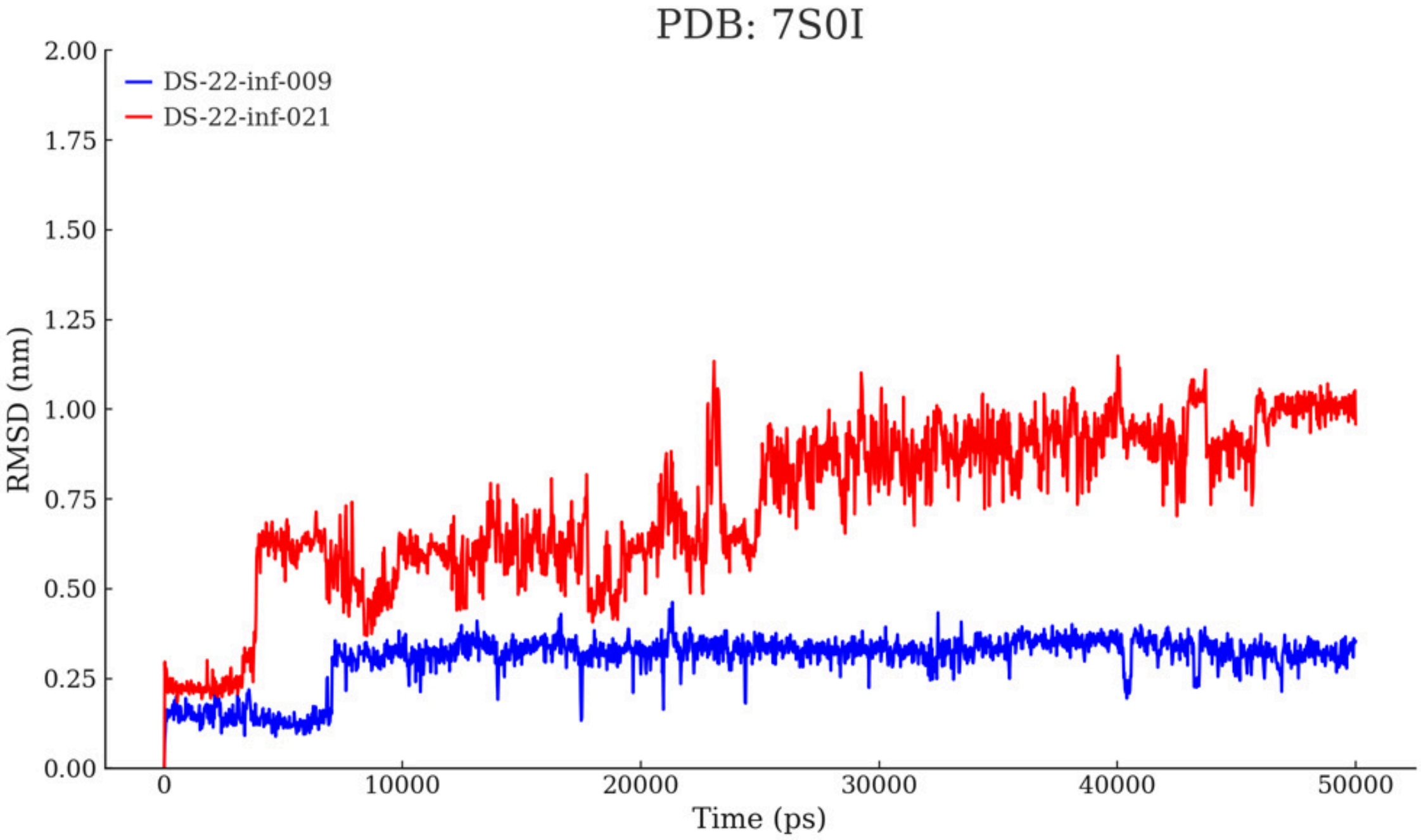

PDB: 7S0I
DS-22-inf-009
DS-22-inf-021
RMSD (nm)
Time (ps)
2.00
1.75
1.50
1.25
1.00
0.75
0.50
0.25
0.00
0
10000
20000
30000
40000
50000

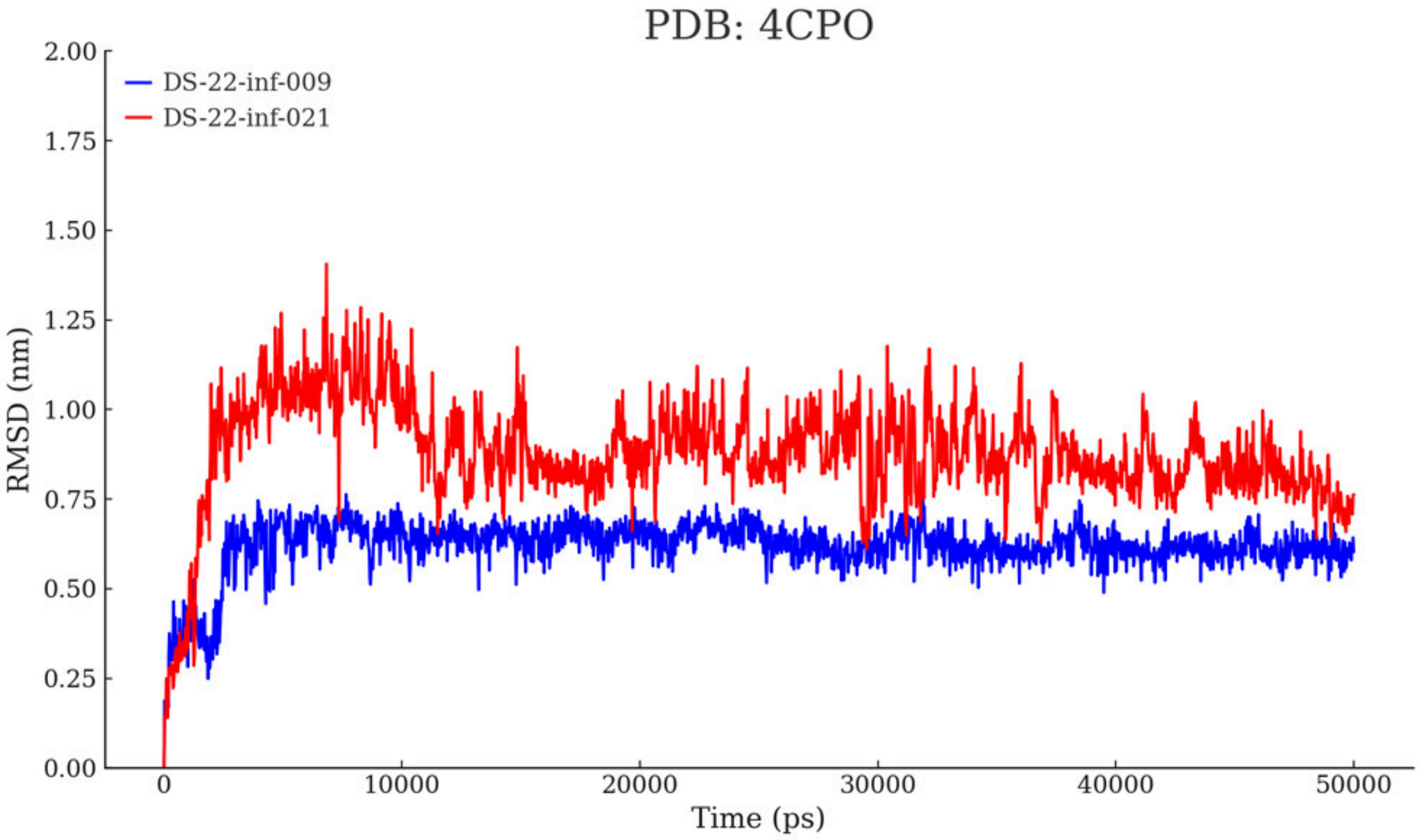

PDB: 4CPO
DS-22-inf-009
DS-22-inf-021
RMSD (nm)
Time (ps)
2.00
1.75
1.50
1.25
1.00
0.75
0.50
0.25
0.00
0
10000
20000
30000
40000
50000

Figure S4

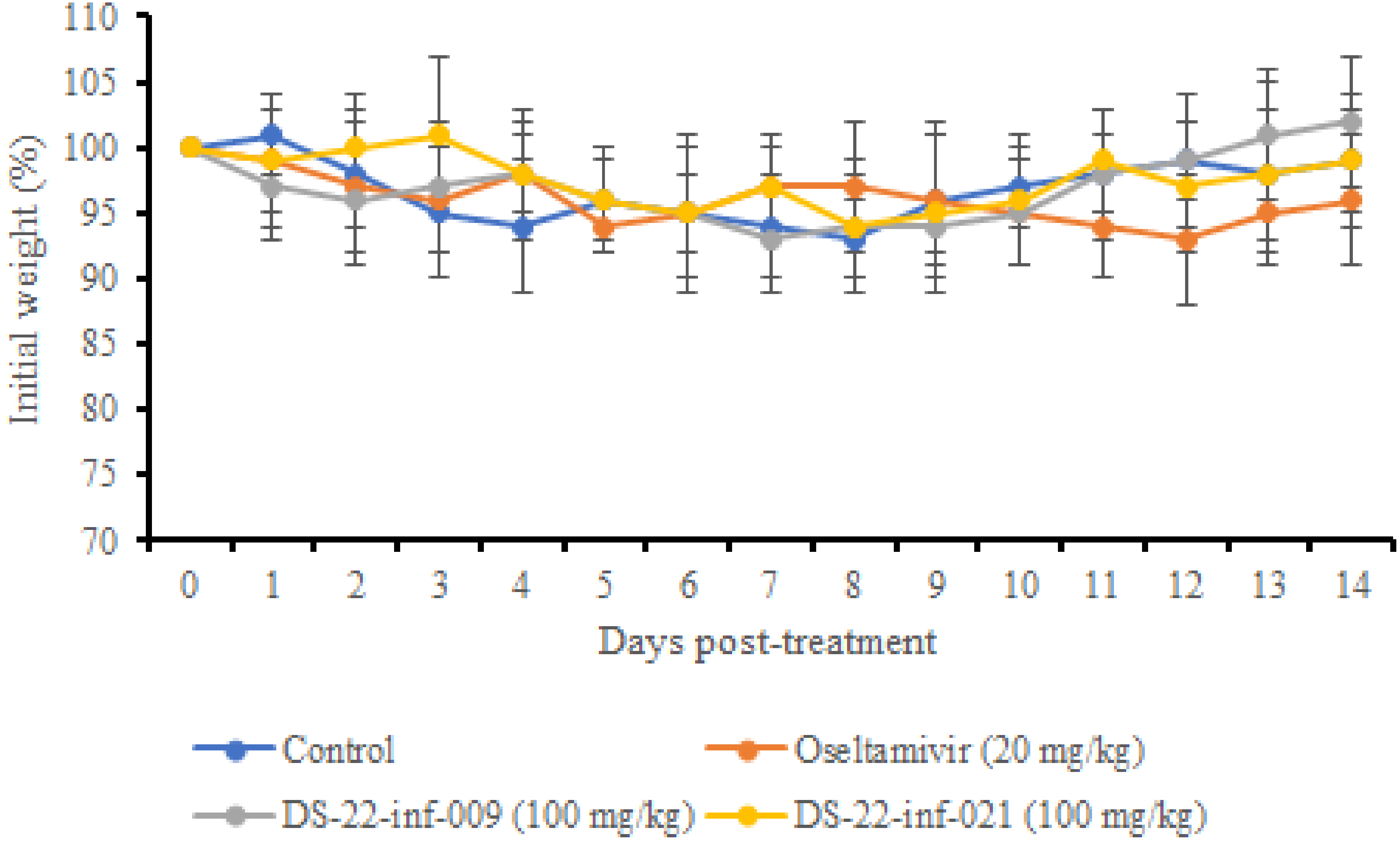
110
105
100
95
90
85
80
75
70
Initial weight (%)
0
1
2
3
4
5
6
7
8
9
10
11
12
13
14
Days post-treatment
Control
Oseltamivir (20 mg/kg)
DS-22-inf-009 (100 mg/kg)
DS-22-inf-021 (100 mg/kg)

**Figure S5**

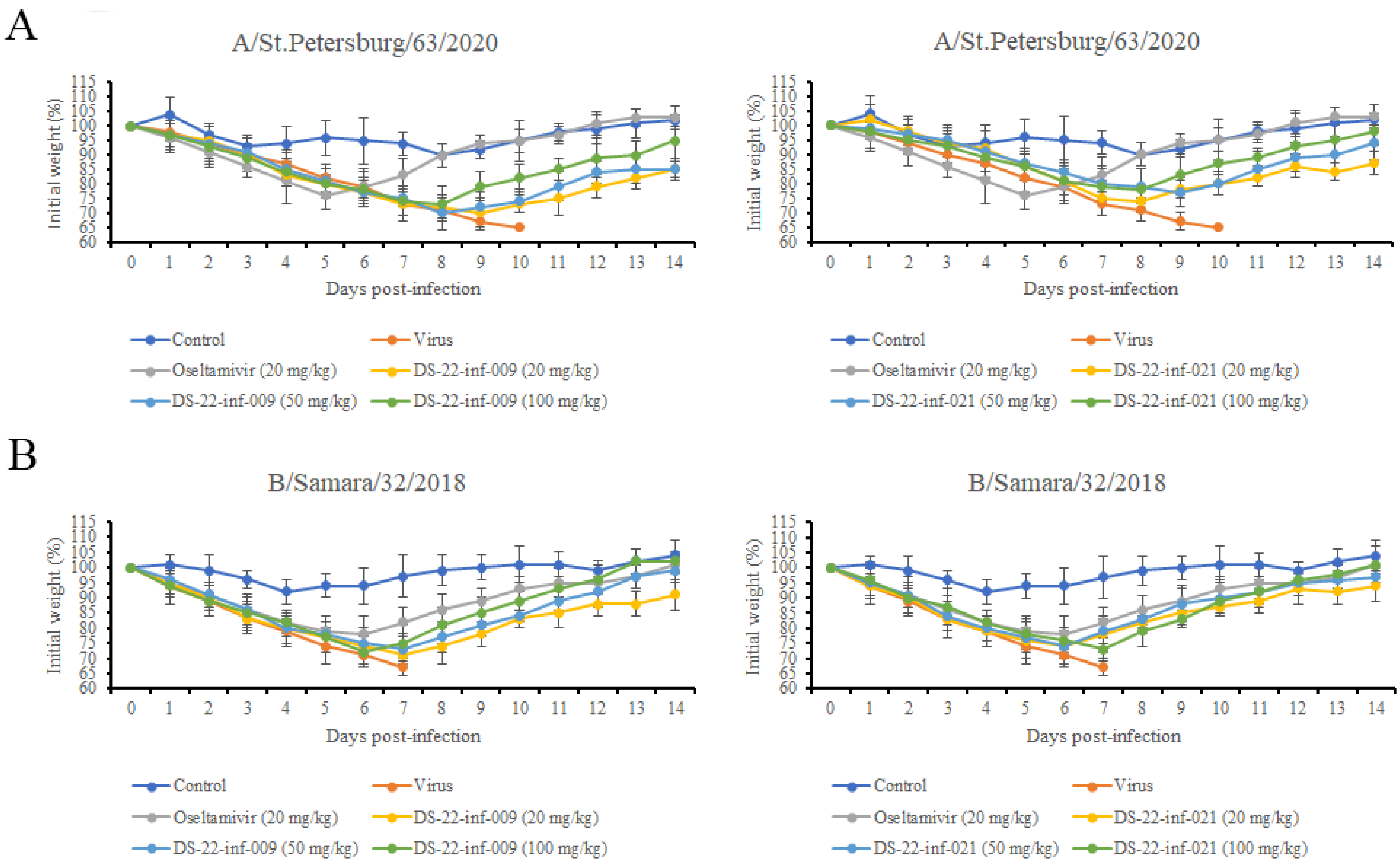
A
A/St.Petersburg/63/2020
Initial weight (%)
Days post-infection
Control
Virus
Oseltamivir (20 mg/kg)
DS-22-inf-009 (20 mg/kg)
DS-22-inf-009 (50 mg/kg)
DS-22-inf-009 (100 mg/kg)
A/St.Petersburg/63/2020
Initial weight (%)
Days post-infection
Control
Virus
Oseltamivir (20 mg/kg)
DS-22-inf-021 (20 mg/kg)
DS-22-inf-021 (50 mg/kg)
DS-22-inf-021 (100 mg/kg)
B
B/Samara/32/2018
Initial weight (%)
Days post-infection
Control
Virus
Oseltamivir (20 mg/kg)
DS-22-inf-009 (20 mg/kg)
DS-22-inf-009 (50 mg/kg)
DS-22-inf-009 (100 mg/kg)
B/Samara/32/2018
Initial weight (%)
Days post-infection
Control
Virus
Oseltamivir (20 mg/kg)
DS-22-inf-021 (20 mg/kg)
DS-22-inf-021 (50 mg/kg)
DS-22-inf-021 (100 mg/kg)